\documentclass[%
 aip,
 amsmath,amssymb,
 reprint,%
]{revtex4-1}

\usepackage{graphicx}
\usepackage{dcolumn}
\usepackage{bm}

\usepackage[utf8]{inputenc}
\usepackage[T1]{fontenc}
\usepackage{mathptmx}
\usepackage{etoolbox}
\usepackage{hyperref}

\makeatletter
\def\@email#1#2{%
 \endgroup
 \patchcmd{\titleblock@produce}
  {\frontmatter@RRAPformat}
  {\frontmatter@RRAPformat{\produce@RRAP{*#1\href{mailto:#2}{#2}}}\frontmatter@RRAPformat}
  {}{}
}%
\makeatother
\begin{document}

\preprint{AIP/123-QED}

\title[]{Coupled Flexural Optomechanical Cavities with Engineered Nanomechanical Interconnects}
\author{David Alonso-Tomás}
\thanks{Corresponding authors: david.alonso@ub.edu \& dnavarro@ub.edu}
\affiliation{MIND-IN2UB, Departament d'Enginyeria Electrónica i Biomédica, Facultat de Física, Universitat de Barcelona, Martí i Franquès 1, Barcelona 08028, Spain}%
\author{Guillermo Arregui}
\affiliation{Swiss Federal Institute of Technology Lausanne (EPFL), CH-1015 Lausanne, Switzerland}%
\author{Bingrui Lu}
\affiliation{Department of Electrical and Photonics Engineering, DTU Electro,
Technical University of Denmark, Building 343, DK-2800 Kgs.\ Lyngby, Denmark.}%
\author{Sergei Lepeshov}
\affiliation{Department of Electrical and Photonics Engineering, DTU Electro,
Technical University of Denmark, Building 343, DK-2800 Kgs.\ Lyngby, Denmark.}
\author{Søren Stobbe}
\affiliation{Department of Electrical and Photonics Engineering, DTU Electro,
Technical University of Denmark, Building 343, DK-2800 Kgs.\ Lyngby, Denmark.}
\affiliation{NanoPhoton - Center for Nanophotonics, Technical University of Denmark,
Ørsteds Plads 345A, DK-2800 Kgs.\ Lyngby, Denmark.}
\author{Daniel Navarro-Urrios}
\thanks{Corresponding authors: david.alonso@ub.edu \& dnavarro@ub.edu}
\affiliation{MIND-IN2UB, Departament d'Enginyeria Electrónica i Biomédica, Facultat de Física, Universitat de Barcelona, Martí i Franquès 1, Barcelona 08028, Spain}%

\date{\today}
             
\begin{abstract}
Integrated nanomechanical circuits require compact and predictable ways to read out, confine, and connect mechanical motion across multiple nanoscale elements. This challenge is particularly acute for megahertz flexural modes, whose large mechanical response and nonlinear dynamics are attractive for optomechanics, sensing, and signal processing, but whose extended nature makes local confinement and coupling difficult within dense devices. Here we demonstrate an optomechanical nanobeam platform in which optical transduction and mechanical connectivity are both engineered lithographically. Transverse geometric asymmetry in the photonic-crystal cavity breaks the cancellation that suppresses dispersive coupling to in-plane flexural motion, making these modes optically bright without ancillary structures. In parallel, serpentine mechanical links engineered through their complex band structure act as compact mirrors and evanescent couplers for MHz flexural waves. In coupled-cavity devices, the normal-mode splitting decays exponentially with the number of serpentine cells, yielding an experimental attenuation constant in quantitative agreement with full-system simulations. Geometry-dependent measurements further show that the coupling can be tuned by the interconnect design and identify regimes where finite-link modes hybridize with the cavity modes, beyond a simple two-resonator picture. These results establish complex-band-engineered mechanical links as calibrated interconnects for scalable optomechanical nanocircuits based on optically addressable MHz flexural resonators.

\end{abstract}
\maketitle

\section{Introduction}   
Nanomechanical resonators are key building blocks for sensing, signal processing, nonlinear dynamics, and hybrid photonic--phononic technologies~\cite{Ekinci2005RSI061101,Xu2022ACSNano15545}. Scaling these capabilities from individual devices to integrated nanomechanical circuits requires more than high-performance resonators: it requires compact means to guide, confine, and couple mechanical motion between neighboring nanoscale elements~\cite{Hatanaka2014NNano520}. In optomechanical implementations, this mechanical connectivity must also be combined with local optical transduction and readout. Optomechanical crystals provide a powerful route toward this goal by co-localizing optical and mechanical modes at the nanoscale, enabling strong light--motion interactions and quantum optical control~\cite{Eichenfield2009Nature08061,Eichenfield2009Nature08524,Aspelmeyer2014RMP1391}. Most implementations have focused on GHz mechanical modes, where phononic mirrors provide strong confinement, resolved-sideband operation, and large single-photon coupling rates~\cite{SafaviNaeini2010OE14926,Chan2011Nature10461,SafaviNaeini2014PRL153603,GomisBresco2014NComms4452,Ren2020NComms3373}.

Megahertz flexural modes offer a complementary regime. Their large driven displacements and low stiffness enable sensitive motion transduction and readily accessible nonlinear dynamics, including coherent phonon generation, chaos, synchronization, and optomechanical frequency-comb dynamics~\cite{Leijssen2015SciRep15974,NavarroUrrios2015SciRep15733,NavarroUrrios2017NComms14965,Colombano2019PRL017402,Wang2024PRL163603}. However, their long effective wavelengths also make them difficult to engineer locally. In contrast to GHz optomechanical-crystal modes, whose confinement and coupling can be controlled with wavelength-scale phononic mirrors, MHz flexural modes are often strongly influenced by the overall device geometry and distant boundary conditions. Dense circuits of individually addressable flexural resonators therefore require compact elements that can both read out in-plane motion and prescribe mechanical coupling within a small footprint.

A first challenge is transduction. In a laterally symmetric nanobeam cavity, in-plane flexural motion only weakly shifts the optical resonance because the perturbations produced on opposite sides of the beam nearly cancel. Stronger coupling can be obtained in zipper, slot-mode, gap-enhanced, and sliced cavities, where the optical field is concentrated across a narrow gap and is highly sensitive to differential motion~\cite{Chan2009OE3802,Davanco2012OE24394,Leijssen2015SciRep15974}. These approaches provide large displacement sensitivity, but they typically rely on an additional optical or mechanical element placed parallel to, or close to, the resonator under test. Scalable flexural-mode circuits would instead benefit from optical readout embedded directly in each nanobeam and sensitive to its own in-plane motion.

A second challenge is mechanical connectivity. Fixed boundaries can define individual MHz flexural resonators by terminating a beam, but they do not provide modular barriers that can be inserted between resonators to prescribe coupling within an integrated network. Conventional phononic-crystal or Bragg shielding can create mechanical stop bands, yet their periods and mirror lengths scale with the long flexural wavelength and can therefore become impractically large for compact MHz circuits. Moreover, phononic shielding, soft clamping, and strain engineering have mainly been exploited to reduce mechanical loss and enhance isolation rather than to provide calibrated inter-resonator coupling~\cite{Tsaturyan2017NNano776,Ghadimi2017NanoLett3501,Ghadimi2018Science764}. A compact, passive, and predictable way to implement embeddable mirrors and tunneling barriers for MHz flexural modes therefore remains needed. Geometry-induced stop bands provide a promising route: serpentine, undulated, curved, and coiled elastic structures can create attenuation regions by folding the propagation path ~\cite{Trainiti2015IJSS260,ZhangParnell2017JAM091007,Willey2022PRApp014035}.

Here we address these two challenges by combining asymmetric photonic-crystal nanobeam cavities with complex-band-engineered serpentine mechanical interconnects. The transverse cavity asymmetry shifts the optical field laterally and makes otherwise weakly transduced in-plane flexural resonances optically bright, without ancillary structures. The serpentine links act as compact MHz mechanical mirrors and evanescent couplers, with their attenuation set by the complex band structure of the unit cell~\cite{Laude2009PRB092301}. Experimentally, we read out MHz flexural motion through the integrated optical cavities, identify the dominant dissipation regime from pressure-dependent mechanical quality factors, and demonstrate coupled-cavity normal-mode splittings that decay exponentially with the number of serpentine cells. These results establish complex-band-engineered mechanical links as calibrated, lithographically defined interconnects for scalable optomechanical nanocircuits in which flexural-mode readout and mechanical coupling are engineered locally by design.

\section{Results}
\subsection{Photonic-crystal cavity design}
\begin{figure*}[t]
    \centering    
    \includegraphics[width =\linewidth]{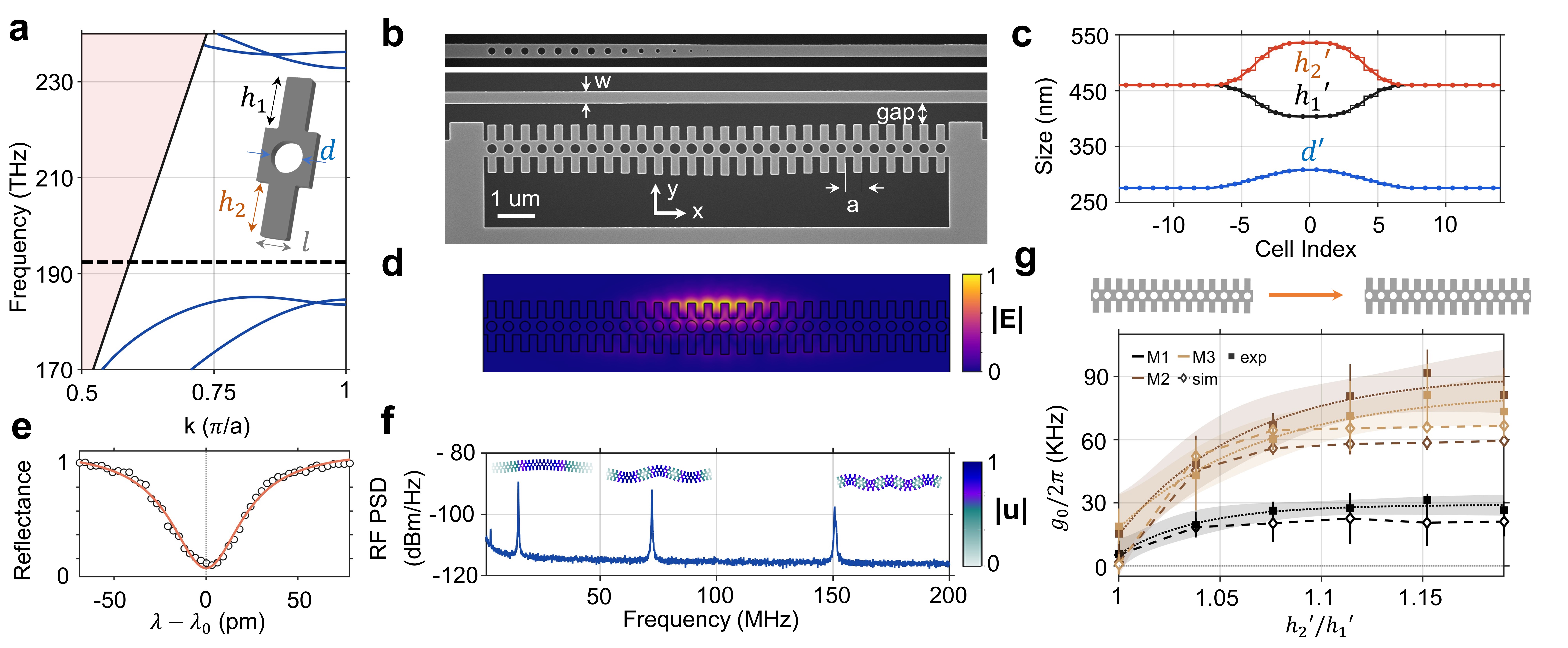}
    \caption{Design, simulation, and characterization of the asymmetric photonic-crystal cavity.
(a) Optical band structure of the mirror unit cell with \((d,h_1,h_2,l)=(276,460,460,230)~\mathrm{nm}\). The red shaded region indicates the radiative region above the light line, while the black dashed line marks the cavity-mode frequency.
(b) SEM image of the fabricated 1D photonic-crystal cavity with a parallel feed-in waveguide. Both configurations are indicated: transmission, without holes in the waveguide, and reflection, with a photonic mirror in the waveguide.
(c) Adiabatic defect profile for the three parameters varied in the unit cell: the two wing dimensions ($h_1'$, $h_2'$) and the diameter of the central hole ($d'$).
(d) Electric-field distribution of the cavity mode.
(e) Optical reflection spectrum around the resonance.
(f) Homodyne radio-frequency signal measured with a spectrum analyzer. The displacement fields of the three dominant mechanical modes are shown, corresponding to in-plane flexural modes with one, three, and five antinodes.
(g) Vacuum optomechanical coupling rate, \(g_0\), as a function of structural asymmetry, defined by the ratio between the two wing dimensions in the defect region, \(h_2'/h_1'\). Experimental error bars correspond to the standard deviation over three nominally identical devices, while simulation error bars are obtained from numerical uncertainty and parameter sensitivity. The dotted curves are saturation fits to the experimental data; shaded regions indicate the corresponding \(1\sigma\) confidence bands.}
    \label{fig1}
\end{figure*}

The first building block of the platform is the optical cavity, implemented as a one-dimensional photonic-crystal nanobeam with corrugated contours derived from the design in Ref.~\cite{GomisBresco2014NComms4452}. Figure~\ref{fig1}a shows the optical band structure of the mirror unit cell, with geometrical parameters \((a,d,h_1,h_2,l)=(460,276,460,460,230)~\mathrm{nm}\) and a thickness of \(220~\mathrm{nm}\). This unit cell opens a photonic bandgap for transverse-electric-like polarization between 185 and \(230~\mathrm{THz}\). The cavity is formed by placing six mirror cells on each side of an adiabatic defect (Fig.~\ref{fig1}b). In the defect region, the lattice constant \(a\) is kept fixed, while \(h_1\), \(h_2\), and the hole diameter \(d\) are smoothly varied over seven transition cells on each side using a cubic-Hermite profile (see Fig.~\ref{fig1}c), reaching a central defect cell with \((h_1',h_2',d')\simeq(406,536,310)~\mathrm{nm}\). The devices were fabricated on a silicon-on-insulator platform using electron-beam lithography, dry etching, and release of the suspended nanobeams; fabrication details are provided in Supplementary Section~S1.\\

The transverse asymmetry of the defect shifts the electric field of the cavity mode toward one side of the nanobeam, as shown in Fig.~\ref{fig1}d. This broken mirror symmetry makes the optical resonance sensitive to in-plane flexural displacement. The cavity is evanescently coupled to a nearby silicon waveguide, Fig.~\ref{fig1}b, designed to match the cavity mode, while the waveguide--cavity gap is adjusted to set the extrinsic decay rate \(\kappa_\mathrm{e}\). The waveguide can be used either in transmission or, by adding a Bragg reflector after the cavity, in reflection. Optical excitation is delivered through a tapered fiber shaped into a microloop~\cite{Ding2010AO2441}, positioned close to a tapered region in the silicon waveguide~\cite{Groeblacher2013APL181104}. The asymmetric cavity design preserves a simulated intrinsic optical quality factor close to \(10^6\). In fabricated devices, reflection spectroscopy, Fig.~\ref{fig1}e, yields a loaded quality factor of \(Q = 31{,}700\), with intrinsic and external decay rates \(\kappa_\mathrm{i}/2\pi = 2.04~\mathrm{GHz}\) and \(\kappa_\mathrm{e}/2\pi = 4.20~\mathrm{GHz}\). Further details of the optical coupling scheme are provided in Supplementary Section~S2.\\

Mechanical spectroscopy is performed using in-fiber balanced homodyne detection; details are provided in Supplementary Section~S3. The spectra show three dominant peaks associated with the one-, three-, and five-antinode in-plane flexural modes (denoted M1, M3 and M5, respectively). Modes with an even number of antinodes are not observed because they exhibit a displacement node at the center of the structure, where the optical mode is localized. Note that shifting the defect along the longitudinal axis of the nanobeam could make these modes optomechanically bright.\\

To understand the role of structural asymmetry, we fabricated three copies of six different geometries with varying \(h_2'/h_1'\) ratios. The simulated optical quality factor slightly decreases as the design approaches the symmetric case, while remaining higher than the experimental values observed for all cavities. We simulated the vacuum optomechanical coupling rate, \(g_0\), by computing the moving-boundary and photoelastic contributions for each geometry~\cite{Johnson2002PRE066611,Eichenfield2009Nature08524}; further details are provided in Supplementary Section~S4.

The experimental values, calibrated using a phase-modulation tone \cite{Gorodetsky2010OE23236}, follow the same trend as the simulations: the vacuum optomechanical coupling rates increase for all modes as the asymmetry is increased, eventually reaching a saturation regime. Although the simulations predict nearly vanishing coupling for fully symmetric structures, the experiments show a non-negligible residual coupling, especially for M3 and M5, which we attribute to fabrication-induced asymmetries. This has allowed previous optomechanical experiments using a symmetric device design \cite{NavarroUrrios2017NComms14965, Colombano2019PRL017402}. Nevertheless, the intentionally asymmetric structures exhibit optomechanical coupling rates in the saturation region that are approximately four to six times larger than those of the symmetric case, with \(g_{0,1}^{a} = 29.0 \pm 1.8~\mathrm{kHz} \approx 5.2\,g_{0,1}^{s}\), \(g_{0,2}^{a} = 91 \pm 8~\mathrm{kHz} \approx 6.0\,g_{0,2}^{s}\), and \(g_{0,3}^{a} = 83 \pm 8~\mathrm{kHz} \approx 4.4\,g_{0,3}^{s}\). This highlights the role of the introduced asymmetry in enabling efficient optical transduction of in-plane flexural modes. We also note that the total optomechanical coupling results from the partial cancellation of two large contributions in the moving-boundary integral, as discussed in Supplementary Section~S4. This suggests that further engineering of the asymmetry could enhance the optomechanical coupling.

\subsection{Complex-band analysis of serpentine interconnects}
\begin{figure*}[t]
    \centering    
    \includegraphics[width =\linewidth]{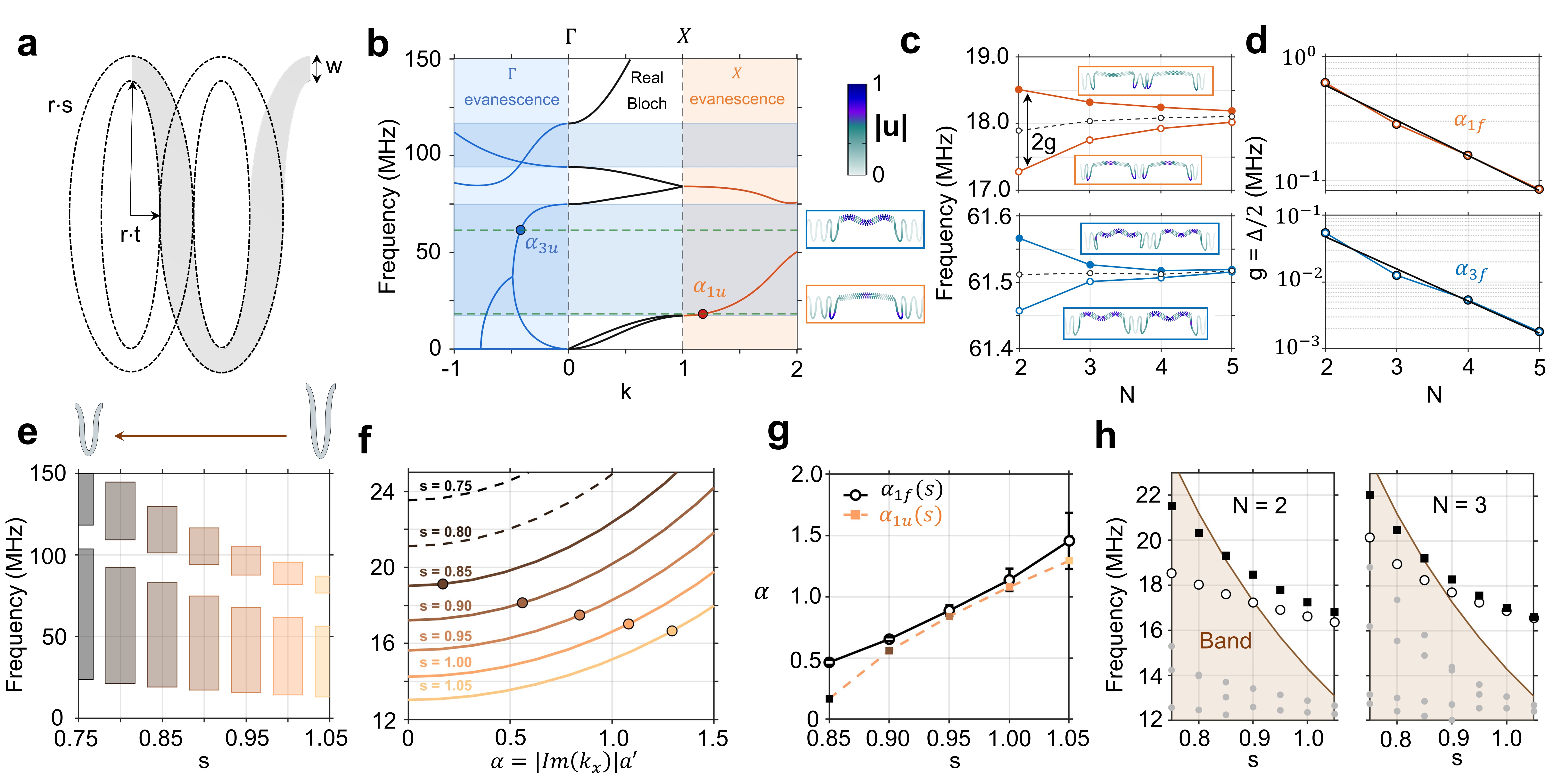}
    \caption{Design and simulation of the mechanical interconnect. (a) Unit cell of the serpentine interconnect. The design can be parametrized by the three parameters of an ellipse. (b) One-dimensional acoustic band structure for the case $(r,s,t)=(2.5~\mu\mathrm{m},0.9,0.15)$. The region between $X$ and $\Gamma$ corresponds to $k_x = \pi k/a' \in(0,\pi/a')$, while the left and right regions account for the continuation of the complex Bloch parameter, $k_x=i k\pi/a'$ and $k_x=\left[1-i(k-1)\right]\pi/a'$, respectively. The gap regions are highlighted in blue, while the green dashed lines indicate the frequencies of the confined in-plane flexural modes, whose displacement fields are shown on the right side of the panel. $\alpha_{3u}$ and $\alpha_{1u}$ are the decay parameters per unit cell, $\mathrm{Im}(k_x)a'$, for the two modes mentioned above. (c) Frequencies of the symmetric and antisymmetric flexural modes when coupled through $N$ cells. Insets show the displacement fields of the coupled geometries. (d) Exponential fit of the mechanical coupling rate $g$, extracted from the splitting between the antisymmetric and symmetric modes, from which $\alpha$ is obtained. (e) Frequency ranges of the band gaps as a function of $s$. (f) Evanescent bands near the $X$-point band edge as a function of $s$. Dots indicate the crossing with the frequency of the one-antinode in-plane flexural mode in the limit $N_\mathrm{mirror}\to\infty$ for each configuration. (g) Comparison between the decay parameter extracted from full-system simulations, $\alpha_{1f}$, following an analysis similar to that in panel (d), and that extracted from the complex band-structure analysis of the unit cell, $\alpha_{1u}$, as a function of $s$. (h) Eigenmode frequencies of the coupled system as a function of \(s\) for \(N = 2\) and \(N = 3\). Small grey dots indicate predominantly serpentine-interconnect eigenmodes, while the brown shaded region indicates the frequency range below the band edge obtained from the Bloch-periodic unit-cell calculation.}
    \label{fig2}
\end{figure*}

Once the nanobeam has been shaped into a photonic-crystal cavity with enhanced sensitivity to in-plane flexural motion, the next step is to design a compact mechanical structure that can confine these modes and mediate evanescent mechanical coupling between neighbouring nanobeams. For this purpose, we use a serpentine interconnect, whose unit cell is shown in Fig.~\ref{fig2}a. The inner contour is defined by an ellipse with horizontal and vertical semi-axes \(r t\) and \(r s\), respectively, where \(r\) sets the overall scale and \(s\) and \(t\) are dimensionless shape parameters. The outer contour is obtained by offsetting this ellipse by the nanobeam width \(w\), so that the interconnect can be attached directly to the clamped nanobeam.

We apply Bloch-periodic boundary conditions along the serpentine period \(a'\), defined as the projected length of one interconnect cell, \(a' = 4rt + 2w\), and compute the one-dimensional mechanical band structure of the unit cell for in-plane modes. Figure~2b shows the result for \((r,s,t) = (2.5~\mu\mathrm{m},0.9,0.15)\), which gives a pitch \(a' = 2.42~\mu\mathrm{m}\). The real Bloch branch is obtained by sweeping the dimensionless parameter \(k\) between the \(\Gamma\) and X points, \(0 \leq k \leq 1\), with \(k_x = k\pi/a'\). This branch reveals two stop bands in which propagating in-plane Bloch modes are absent and the response is evanescent. To quantify the attenuation inside these gaps, we continue the Bloch wavevector into the complex plane on both sides of the real branch: from the \(\Gamma\) point using \(k_x = i k\pi/a'\) for \(-1 \leq k < 0\), and from the X point using \(k_x = [1-i(k-1)]\pi/a'\) for \(1 < k \leq 2\). In this representation, the imaginary part of the Bloch wavevector gives the attenuation per unit cell, \(\alpha = \mathrm{Im}(k_x)a'\), experienced by an in-plane flexural wave at a given frequency inside the gap.

Using FEM simulations of isolated nanobeams terminated by long serpentine mirrors, we estimate the frequencies of the one- and three-antinode in-plane flexural modes. The intersections of these frequencies with the corresponding complex bands provide the unit-cell attenuation constants predicted from the interconnect alone, \(\alpha_{1u}=0.55\) and \(\alpha_{3u}=1.32\). We then test this prediction in a full coupled-cavity geometry, consisting of two nanobeams connected by \(N\) serpentine cells and confined on each outer side by two cells with \(s = 1\). As discussed below, increasing \(s\) increases the attenuation constant per cell.  This system can be described as two coupled identical mechanical resonators with coupling rate \(g\), giving rise to symmetric and antisymmetric hybridized modes separated by \(2g\). Since the evanescent tail decays exponentially along the interconnect, the coupling is expected to follow \(g(N)=g(0)e^{-N\alpha}\). Figure~\ref{fig2}c shows the simulated frequencies of the symmetric and antisymmetric modes as a function of \(N\) for the two in-plane flexural modes considered here. As \(N\) increases, the splitting decreases and the two hybridized frequencies converge towards the isolated-cavity values. We extract \(g\) from half of the frequency splitting and fit its decay with \(N\), obtaining \(\alpha_{1f}=0.65\pm0.03\) and \(\alpha_{3f}=1.11\pm0.08\) for the parameters of Fig.~\ref{fig2}b. The coupling therefore follows the expected exponential behaviour, and the extracted attenuation constants are close to those predicted from the unit-cell complex band structure. The remaining differences are expected because the unit-cell calculation describes an ideal infinite interconnect, whereas the full coupled system includes finite mirrors and cavity--interconnect impedance mismatch. \\

After establishing the exponential dependence on the number of interconnect cells, we study how the geometrical parameters of the serpentine modify its band structure. The full parameter sweep is discussed in Supplementary Section~S5; here we focus on the parameter \(s\), which controls the vertical semi-axis of the inner ellipse, while
keeping the other geometrical parameters fixed \((r,t) = (2.5~\mu\mathrm{m},0.15)\). As \(s\) decreases, the stop-band regions shift to higher frequencies, as shown in Fig.~\ref{fig2}e. We use this dependence to explore whether the attenuation can be tuned continuously by modifying the unit cell itself, rather than only by changing the number of cells. Figure~\ref{fig2}f shows the complex-band continuation near the \(X\) point for \(s\) ranging from 0.75 to 1.05. For each value of \(s\), we indicate the crossing between the complex band and the frequency of the isolated one-antinode flexural mode, M1, calculated for that same geometry. This is important because changing \(s\) not only modifies the evanescent decay of the interconnect, but also changes the boundary condition seen by the nanobeam and therefore shifts the frequency of the confined mode. For \(s\lesssim0.85\), the M1 frequency moves out of the stop band, and the simple evanescent-coupling picture is expected to break down. To quantify this behaviour, we repeat the analysis of Figs.~\ref{fig2}c and \ref{fig2}d for each value of \(s\), fitting the decay of the coupling rate with \(N\) to extract \(\alpha_{1f}(s)\). Figure~\ref{fig2}g compares this value with the attenuation predicted from the unit-cell complex band structure, \(\alpha_{1u}(s)\). Both quantities show a clear decrease as \(s\) is reduced towards the band edge, confirming that the coupling can be tuned through the geometry of the interconnect. Close to \(s\simeq0.85\), however, the fitted attenuation starts to deviate from the unit-cell prediction. In this regime, the confined mode lies near the band edge, and the coupling is no longer governed solely by a single evanescent tail.

This breakdown is clarified in Fig.~\ref{fig2}h, where we plot the symmetric and antisymmetric coupled-cavity modes together with the modes associated with the finite serpentine interconnect as a function of \(s\), for \(N=2\) and \(N=3\). For short interconnects, the finite-link modes remain sufficiently detuned and do not strongly perturb the symmetric and antisymmetric doublet. As \(N\) increases, however, the spectrum of the finite serpentine approaches the band structure of the infinite interconnect, allowing link modes to hybridize with the coupled-cavity modes. This hybridization introduces additional splittings that are not captured by the simple evanescent-coupling model between two isolated flexural resonators.

\subsection{Experimental characterization of serpentine clamped flexural resonances}

\begin{figure*}[t]
    \centering    
    \includegraphics[width =\linewidth]{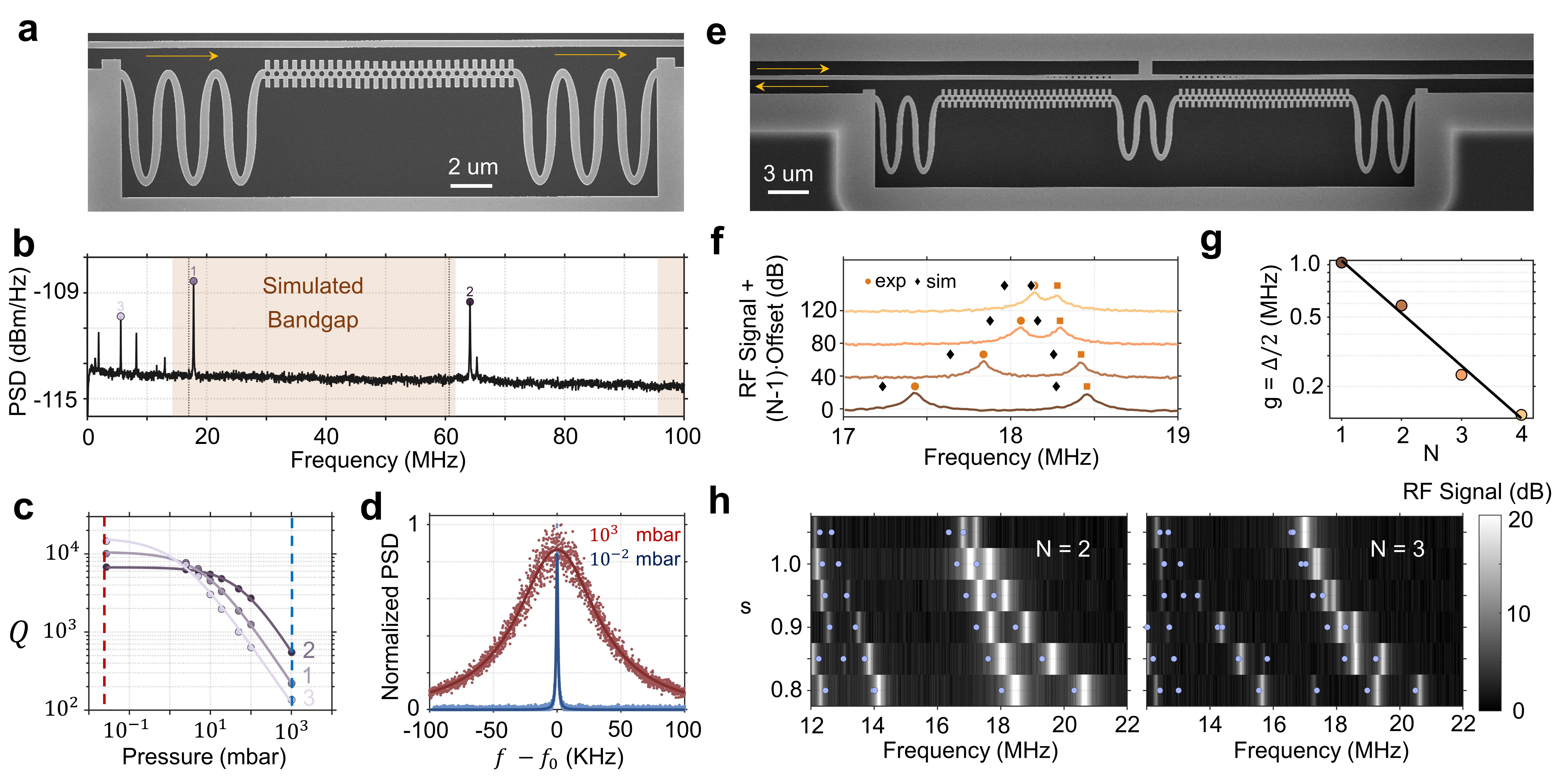}
    \caption{Characterization of the confined and mechanically coupled nanobeams. (a) SEM image of a single nanobeam PhC clamped using the unit cell described in Fig.~2. (b) RF spectra of the mechanical modes. The three dominant peaks are labeled as 1, 2, and 3. The brown region indicates the simulated band gap for this particular set of parameters, $(r,s,t)=(2.5~\mu\mathrm{m},1,0.15)$, while the dashed lines indicate the simulated frequencies of modes 1 and 2. (c) Quality factor of the highlighted modes as a function of chamber pressure. (d) Comparison of the mechanical Lorentzian peak at atmospheric pressure and in the saturated-\(Q\) regime. (e) SEM image of the mechanically coupled nanobeams. Two Bragg mirrors in the waveguide allow each cavity to be probed independently. (f) RF signal detected in one of the optical cavities as a function of \(N\). (g) Exponential fit of the mechanical coupling as a function of \(N\). (h) Contour RF plot of the mechanical spectra as a function of \(s\) for \(N=2\) and \(N=3\). Blue dots indicate the simulated values.}
    \label{fig3}
\end{figure*}

We first characterize a single nanobeam cavity terminated by serpentine mirrors. In the present work, the primary role of these mirrors is to place the flexural resonances within an engineered stop band and to provide the evanescent barriers used below to control inter-cavity coupling. The same stop band is designed to suppress the radiation of in-plane flexural waves into the supports. Figure~\ref{fig3}a shows an SEM image of a device with \((r,s,t)=(2.5~\mu\mathrm{m},1,0.15)\) and \(N=3\), for which the complex-band analysis predicts an attenuation of approximately \(10~\mathrm{dB}\) per cell at the frequency of the simulated fundamental in-plane flexural mode. According to FEM simulations, this attenuation is predicted to support clamping-loss-limited quality factors above \(10^9\) for M1, as discussed in Supplementary Section~S6.

We measure the device in a cryostat that allows the pressure to be varied from \(100~\mathrm{mbar}\) down to \(10^{-5}~\mathrm{mbar}\). The optical cavity is probed in transmission using grating couplers~\cite{Hansen2023OE17424}, and the RF signal is acquired by direct detection. Figure~\ref{fig3}b shows the corresponding RF spectrum arising from optomechanical transduction of the cavity mode; further details are provided in Supplementary Section~S3. The dashed lines indicate the simulated frequencies of the localized flexural modes, while the shaded region marks the simulated stop band of the serpentine mirror. The measured mechanical resonances are slightly blueshifted relative to simulations, especially at higher frequencies, as observed for the three-antinode flexural mode. This trend may arise from small fabrication-induced deviations of the nanobeam geometry, which can affect different flexural modes unequally. By contrast, a uniform residual tensile stress would be expected to shift the mechanical spectrum more uniformly and therefore does not fully account for the observed frequency-dependent discrepancy. At lower frequencies, additional peaks associated with finite-serpentine modes are also observed, such as the mode labelled 3. These modes exhibit weaker transduction than the nanobeam modes because their motion produces only a small transverse displacement of the optical cavity region.

Figure~\ref{fig3}c shows the pressure dependence of the quality factors of the three dominant mechanical resonances. At high pressure, all modes are limited by gas damping, as discussed in Supplementary Section~S7. Upon evacuation, the quality factors increase and eventually saturate at values of order \(10^4\). The highest saturated quality factor is obtained for mode 3, which is mainly associated with the serpentine mirror itself. This indicates that acoustic radiation through the supports is not the dominant loss channel in the present devices. Consistently, FEM simulations predict much higher clamping-loss-limited quality factors, even for nanobeams without serpentine terminations, than the saturated values measured experimentally. We therefore attribute the observed saturation to internal dissipation mechanisms, such as thermoelastic damping, surface loss, or material loss in the released silicon structure~\cite{Lifshitz2000PRB,Yasumura2000JMEMS,Ghaffari2013SciRep}. This interpretation is further supported by low-temperature measurements at \(4~\mathrm{K}\), which show an increase in the quality factor for all modes, as reported in Supplementary Section~S7. The nearly linear pressure dependence of the linewidth between \(10^3\) and \(1~\mathrm{mbar}\), Fig.~\ref{fig3}d, also suggests possible compact on-chip pressure-sensing operation in the gas-damping regime~\cite{Chen2022ACSAMI,Reinhardt2024ACSPhotonics}.

\subsection{Experimental characterization of flexural mode molecules}
Although the single-nanobeam measurements identify flexural resonances consistent with localized modes, they do not by themselves prove the exponential attenuation predicted from the complex band structure. To test this prediction experimentally, we implement the coupled-cavity geometry discussed in the previous section, where two nanobeams are connected by \(N\) serpentine cells. Figure~\ref{fig3}e shows an SEM image of one of the fabricated devices, with \((N,r,s,t)=(2,2.5~\mu\mathrm{m},0.9,0.15)\) and 2 cells with s = 1 on each side. The device includes two adiabatic Bragg reflectors in the access waveguide, allowing the two cavities to be probed independently in reflection. In the following, we focus on the response of a single cavity; a comparison between both cavities is provided in Supplementary Section~S8.

We first repeat the analysis of Figs.~\ref{fig2}c and \ref{fig2}d. We perform optomechanical spectroscopy on four devices with \(N=1\) to \(4\), probing one of the photonic-crystal cavities in each case, and compare the measured spectra with the simulated eigenfrequencies, as shown in Fig.~\ref{fig3}f. As in the single-nanobeam devices, the experimental resonances are slightly shifted toward higher frequencies. Nevertheless, the symmetric and antisymmetric modes follow the same overall trend predicted by the simulations. Extracting the coupling rate \(g\) from half of the measured splitting, which is a good approximation given that the experimental detuning of the intrinsic frequencies is smaller than the observed coupling (see Supplementary Information S8 for more details), and fitting its dependence on \(N\), as shown in Fig.~\ref{fig3}g, gives \(\alpha_{\mathrm{exp},1}=0.69\pm0.05\), in agreement with the simulated full-system value \(\alpha_{1f} = 0.65 \pm 0.03\). This experimentally confirms that the serpentine interconnect acts as a compact evanescent barrier for in-plane flexural modes, with an attenuation that can be predicted and engineered through its geometry.

Finally, Fig.~\ref{fig3}h shows the experimentally measured dependence of the coupled modes on the serpentine parameter \(s\). The measured spectra remain close to the simulated eigenfrequencies and reproduce the finite-mirror effect discussed above: for \(N=2\), the finite-serpentine modes remain sufficiently detuned and have little influence on the symmetric--antisymmetric splitting, whereas for larger \(N\), modes of the interconnect itself can hybridize with the flexural cavity modes. These measurements confirm that the coupled-cavity spectra are governed not only by the evanescent decay inside the stop band, but also by the finite spectrum of the serpentine link when the interconnect becomes longer or the mode approaches a band edge.

\section{Conclusion}
In conclusion, we have demonstrated an optomechanical platform for MHz in-plane flexural modes that combines asymmetric photonic-crystal nanobeam cavities with complex-band-engineered serpentine interconnects. The transverse cavity asymmetry provides a design-controlled dispersive coupling to in-plane motion, converting a symmetry-suppressed and fabrication-sensitive readout channel into a deterministic one; in the present fabrication run, this yields measured vacuum optomechanical coupling rates four to six times larger than in nominally symmetric structures. The serpentine links act as compact mechanical mirrors and evanescent couplers, with an attenuation length set by the complex band structure of the unit cell. This framework quantitatively captures the exponential decay of the mechanical coupling with interconnect length, providing a direct design rule for coupled flexural resonators.

Although the present devices are not limited by engineered clamping loss, pressure- and temperature-dependent measurements show that internal dissipation dominates the measured quality factors after evacuation. With improved material, surface, and stress engineering, the same stop-band design could allow acoustic leakage suppression to become directly observable, as in GHz optomechanical-crystal resonators embedded in acoustic-bandgap shields and operated at millikelvin temperatures~\cite{Meenehan2014PRA,MacCabe2020Science}.

Because the relevant design parameter is the complex band structure of the interconnect, this procedure is generic and can be transferred to other material platforms, frequency ranges, and optical transduction schemes. Within the silicon nanobeam platform demonstrated here, larger arrays, rings, and lattices based on these links could provide compact building blocks for phonon-routing, coupled-cavity, topological, and sensing architectures~\cite{Fang2016NatPhoton,Madiot2023PRL,Bereyhi2022PRX,Stassi2019NatCommun}. By enabling controlled mechanical coupling between multiple optically readable flexural resonators, this platform could also be used to explore multimode nonlinear dynamics, including intermodal coupling, synchronization, and multimode phonon lasing~\cite{Asadi2018PTRSA,AlonsoTomas2026LPR}. More broadly, complex-band-engineered mechanical links could provide a passive connectivity layer that complements optically programmable interactions, offering a route toward hybrid optomechanical lattices with both engineered static coupling and reconfigurable dynamics~\cite{Slim2024Nature}.

\section{Acknowledgments}
This work was supported by the MICIU through the projects ALLEGRO (Grant No. PID2021-124618NB-C22) and OMENS (Grant No. PID2024-156058NB-I00). D.A.T acknowledges a FPU fellowship (Grant No. FPU23/00690). B.L., S.L., and S.S. gratefully acknowledge financial support from the Innovation Fund Denmark (Grants No.\ 4356-00007B -- EQUAL), the European Research Council (Grant No.\ 101045396 -- SPOTLIGHT), the Horizon Europe Research and Innovation Programme (Grant No.\ 101098961 -- NEUROPIC, and the Danish National Research Foundation (Grant No.\ DNRF147 -- NanoPhoton).

\section{Disclosures}
The authors declare no conflicts of interest

\section{Data Availability}
The data that support the findings of this study are available from the corresponding author upon reasonable request.

\section{Supplementary Material}

\section*{S1. Fabrication}
The devices are patterned into the 220 nm silicon device layer of a silicon-on-insulator (SOI) chip, which also includes a 2 µm buried oxide layer and a 775 µm silicon handle layer. We fabricate the devices using electron-beam lithography, plasma reactive-ion etching, and selective underetching. First, electron beam lithography is used to pattern a 180 nm ZEP 530A6 resist with 0.4 nA current and a shot pitch of 1 nm. Following development, we use reactive-ion etching to transfer the pattern in the resist to a 30 nm Cr hard mask layer via a 12 nm intermediate amorphous silicon layer, both deposited by an electron-beam evaporator on top of the silicon device layer. Without breaking the vacuum, we continue to etch the silicon device layer using the Cr etch mask to produce high-aspect ration nanostructures with vertical sidewalls. Finally, vapor-phase HF etching removes the buried oxide beneath the devices, releasing the cavity, grating couplers, and waveguides into suspended structures attached to the rest of the chip by tethers and springs.

\section*{S2. Waveguide design}
Here, we describe the design of the parallel waveguide used to couple light into the photonic crystal cavities.

\subsection*{Mode matching}

\begin{figure}[b]
\centering    
\includegraphics[width= \linewidth]{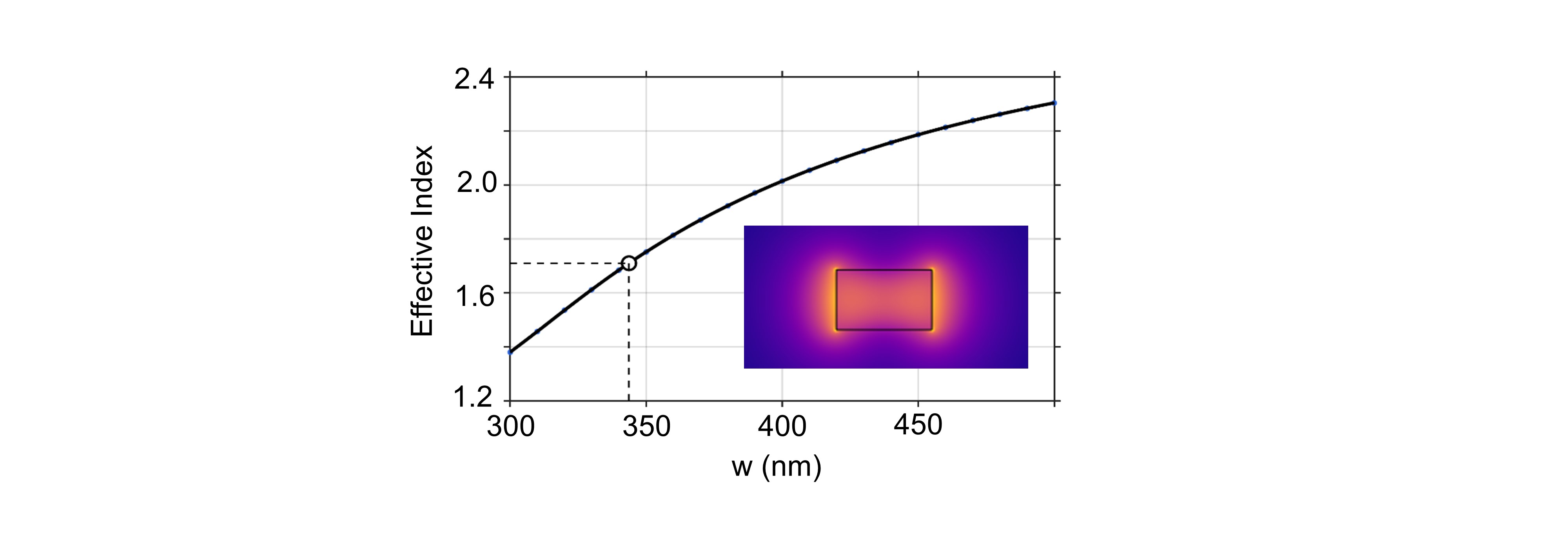}
\caption{Effective index of the fundamental TE-like optical mode as a function of the waveguide width. The inset shows the normalized electric-field distribution of the simulated mode. \label{fig: matching}}
\end{figure}
The design procedure starts by computing the effective index of the fundamental transverse-electric-like mode of the access waveguide as a function of its width. This calculation is performed in COMSOL Multiphysics using a cross section of the waveguide. Since the cavity is designed with a fixed pitch '\(a_c\)', the Bloch effective index associated with the defect mode at the $X$ point can be estimated as
\begin{equation}
n_{\mathrm{eff}}^{\mathrm{Bloch}} = \frac{\lambda_0}{2a_c} \approx 1.7,
\end{equation}
using the simulated cavity resonance frequency of $192.1~\mathrm{THz}$. We then use the calculated waveguide effective-index curve to choose a waveguide width that approximately matches this value. This procedure gives a suitable working width of about $340~\mathrm{nm}$, as shown in Fig.~\ref{fig: matching}.

\subsection*{Mirror}

\begin{figure*}[t]
\centering    
\includegraphics[width= 0.8\linewidth]{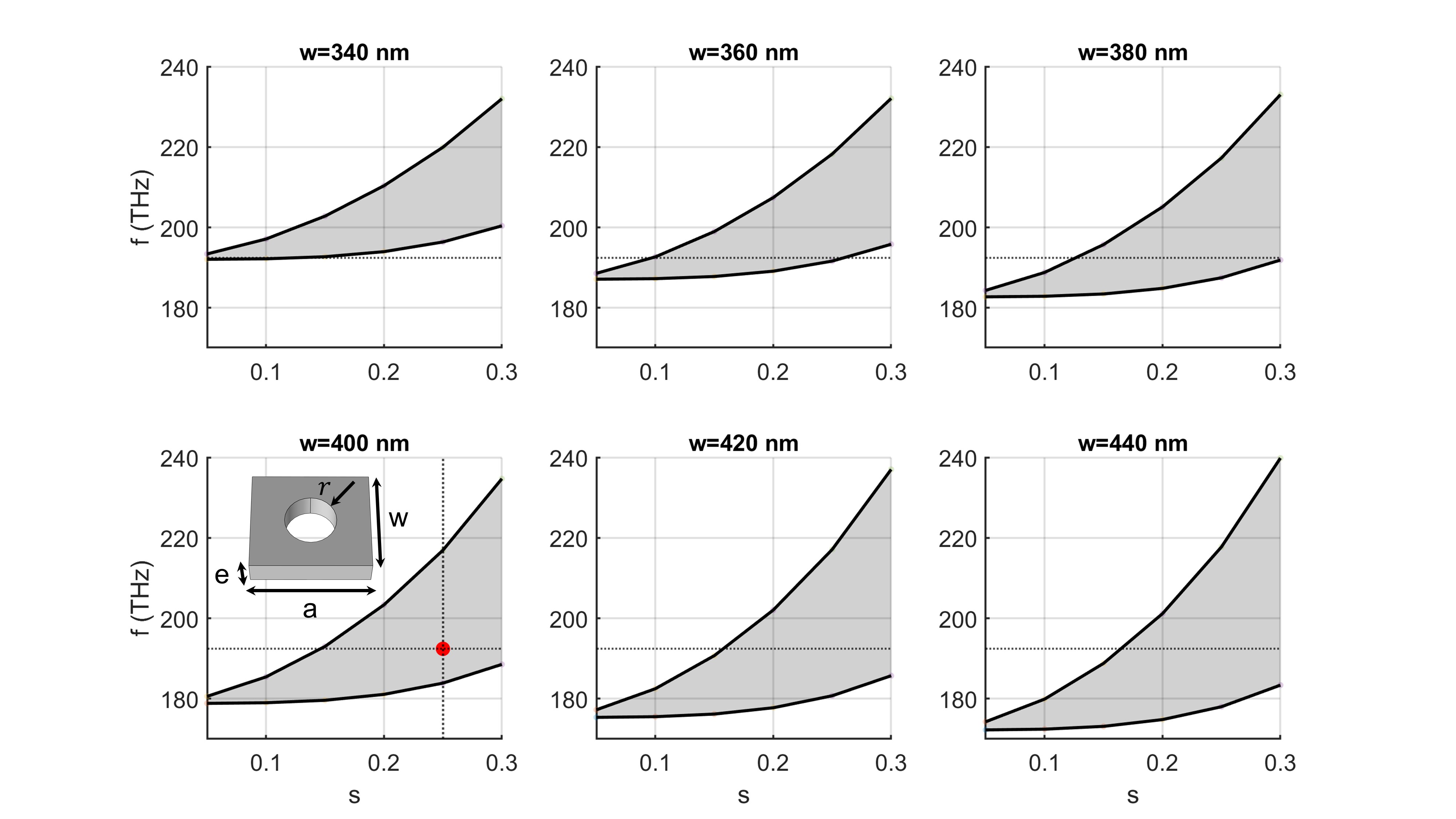}
\caption{ Bandgaps obtained from the band structure of the unit cell (shown in the inset) for different values of the parameter $s=r/w$ and different waveguide widths $w$.\label{fig: mirrorvar}}
\end{figure*}

The photonic-crystal mirror is designed by performing an optical eigenfrequency study and real band-structure analysis of a unit cell. The unit cell consists of a silicon waveguide section with pitch $a = a_c$, width $w$, thickness $e$, and a central hole of radius $r$. \\

Figure~\ref{fig: mirrorvar} shows the stop-band regions obtained as a function of $s = r/w$ for different waveguide widths. We find that keeping the mirror width equal to that of the access waveguide does not open a bandgap at the cavity-mode frequency. Therefore, we choose $w = 400~\mathrm{nm}$ and $s = 0.25$, placing the cavity mode inside the bandgap while keeping the mirror geometry close enough to the access waveguide to allow a low-loss adiabatic transition. \\

\begin{figure*}[t]
\centering    
\includegraphics[width= 0.8\linewidth]{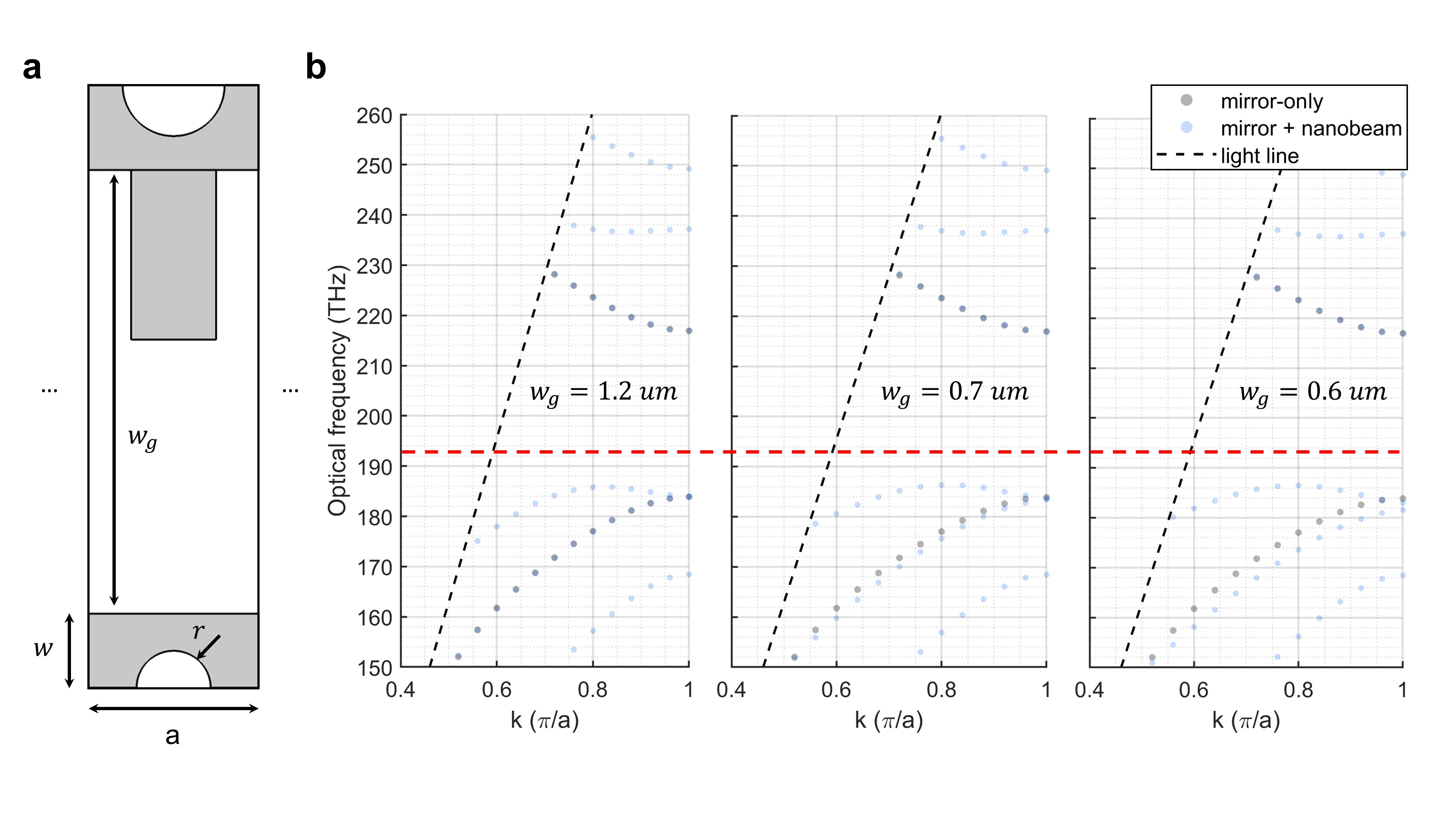}
\caption{ \textbf{(a)} Unit cell of the complete system, including the waveguide mirror and the photonic-crystal cavity. \textbf{(b)} Band structure of the unit cell for different values of the gap $w_g$ between the waveguide and the nanobeam. \label{fig: mirrorgap}}
\end{figure*}

After this analysis, we also study how the presence of a nearby nanobeam at a distance $w_g$ may modify the bandgap properties. We find that the band structure is noticeably affected only for separations below $700~\mathrm{nm}$ (see Fig. ~\ref{fig: mirrorgap}). Since the gap used in the experiments is always larger than $1100~\mathrm{nm}$, the dispersive effect of the feed waveguide on the cavity can be neglected.

\section*{S3 Experimental setups and coupling configurations}
\subsection*{Experimental setups}
In the experiments discussed in the main text, we use two different schemes to perform in-fiber optomechanical spectroscopy.

\begin{figure*}[t]
\centering    
\includegraphics[width= 0.8\linewidth]{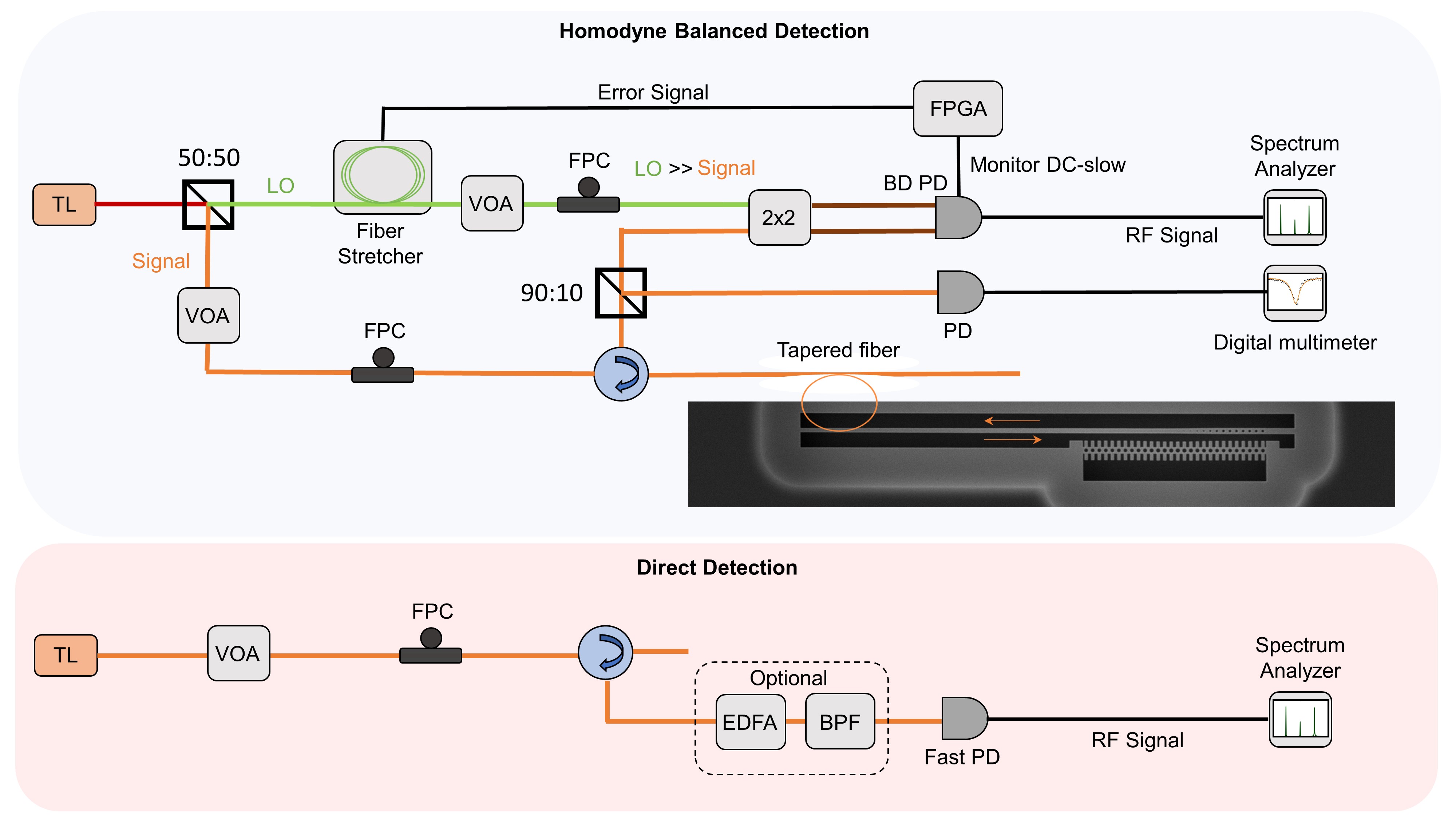}
\caption{Both experimental implementations used in this work. TL: tunable laser; VOA: variable optical attenuator; FPC: fiber polarization controller; LO: local oscillator; PD: photodetector; EDFA: erbium-doped fiber amplifier; BPF: band-pass filter. The inset shows an SEM image of the coupling region between the tapered fiber and the cavity waveguide.  \label{fig: experimentalsetup}}
\end{figure*}

The simplest approach is direct detection. In this scheme, a tunable laser is coupled to an optical fiber and sends coherent light to the sample. The input power is controlled with a variable optical attenuator (VOA), while the polarization is adjusted using a fiber polarization controller (FPC). The measurement can be performed either in reflection or transmission. A reflection configuration is illustrated in Fig.~\ref{fig: experimentalsetup}. In this case, the light reflected from the sample can be optionally amplified with an EDFA and filtered with a band-pass filter (BPF) to reduce the amplified spontaneous emission noise. The signal is then detected with a fast photodetector (New Focus 1544-B, 12 GHz bandwidth), and the resulting RF signal is measured with a spectrum analyzer (MS2830A).

The main limitation of this scheme is that, even after spectral filtering, the EDFA increases the noise background and can reduce the measurement sensitivity. Conversely, operating without an EDFA requires increasing the input optical power so that enough light reaches the fast photodetector. Due to the bandwidth and sensitivity characteristics of the detector, this typically requires optical powers of several $\mu\mathrm{W}$ at the detector input, which can correspond to tens or hundreds of $\mu\mathrm{W}$ at the cavity input, depending on the coupling efficiency and optical losses in the setup. Such powers can influence the dynamics of the mechanical resonators through optical backaction or direct heating.

An alternative and widely used approach is balanced homodyne detection. In this scheme, the laser light is split into a local oscillator (LO), which does not interact with the sample, and a signal arm, which is sent to the device. The reflected or transmitted signal is then interfered with the LO in a Mach--Zehnder interferometer, and the two outputs are detected with a balanced photodetector. Common-mode noise is subtracted electronically, allowing the measurement to approach the shot-noise limit.

Our implementation, shown in Fig.~\ref{fig: experimentalsetup}, is fully fiber-based. The phase difference between the LO and signal arms is stabilized using a feedback loop implemented with an FPGA (Red Pitaya STEMlab 125-14) and a fiber stretcher driven by a piezoelectric actuator. Light is detected in a fiber-coupled balanced photodetector (PDB48xC). This allows us to measure a selected optical quadrature and improve the detection sensitivity while using very low input powers, below $1~\mu\mathrm{W}$. \\

\subsection*{Coupling configurations}
The experimental setups described above are general schemes for analyzing the optical signal originating from the cavity. However, different approaches can be used to couple light between the integrated waveguide and an optical fiber.

\begin{figure*}[t]
\centering    
\includegraphics[width= 0.8\linewidth]{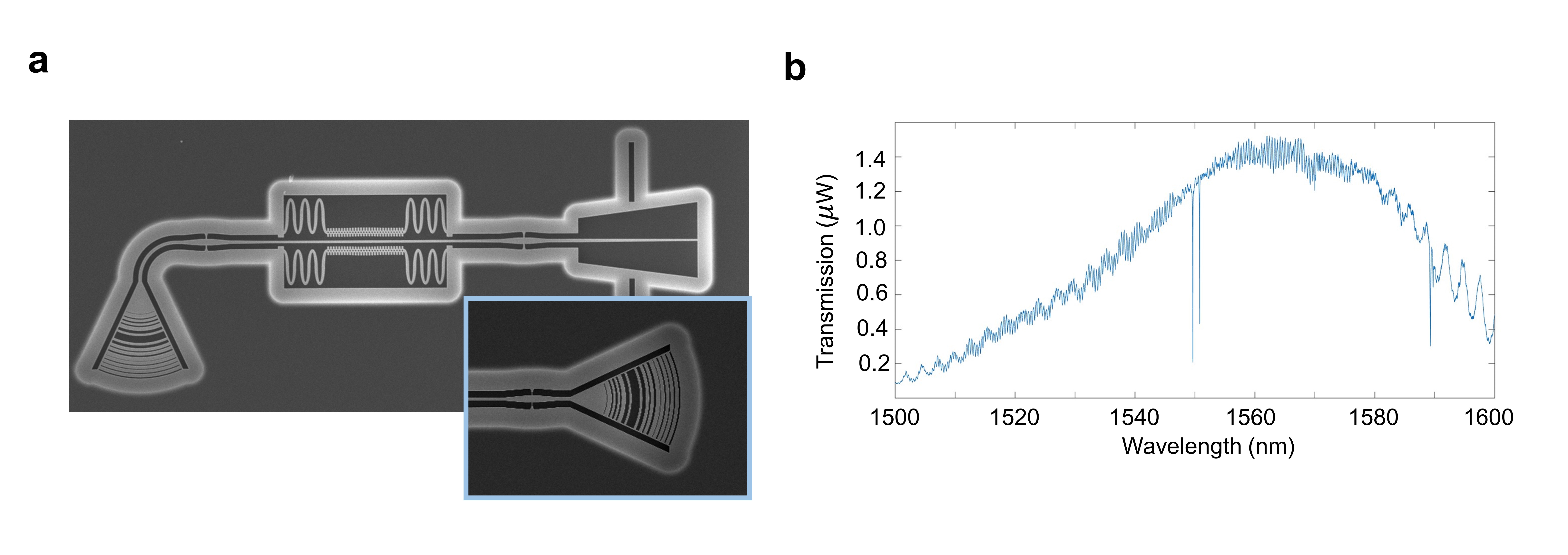}
\caption{\textbf{a} SEM image of single-nanobeam serpentine-clamped structures, combining two input/output coupling schemes. The inset shows the relative orientation of the grating couplers when two of them are used. \textbf{b} Optical transmission spectrum of the grating--grating configuration shown in \textbf{a}.\label{fig: sem}}
\end{figure*}

In most of the experiments described in the main text, we use a tapered optical fiber shaped into a microloop \cite{Ding2010AO2441}. For cryogenic measurements, however, we employ grating couplers. Figure~\ref{fig: sem}a shows an example of a device implementation combining both approaches: a grating coupler and a tapered waveguide. Alternatively, two grating couplers rotated by $90^\circ$ can be used for optical input and output with orthogonal polarizations, which can subsequently be separated using a polarizing beam splitter. Further details on this configuration can be found in \cite{Hansen2023OE17424}.

An example optical spectrum obtained using the latter configuration is shown in Fig.~\ref{fig: sem}b. The broad spectral envelope corresponds to the bandwidth of the grating coupler, while the cavity resonances are observed around $1550~\mathrm{nm}$.

\section*{S4 Optomechanical coupling in asymmetric cavities}

\subsection*{Simulation}
The vacuum optomechanical coupling rate is computed as the optical frequency shift induced by the zero-point motion of the mechanical mode,
\begin{equation}
\frac{g_0}{2\pi}
=
x_{\mathrm{zpf}}
\frac{\partial \nu_{\mathrm{c}}}{\partial x},
\end{equation}
where \(\nu_{\mathrm{c}}\) is the optical resonance frequency and \(x_{\mathrm{zpf}}\) is the zero-point fluctuation amplitude,
\begin{equation}
x_{\mathrm{zpf}} =
\sqrt{\frac{\hbar}{2m_{\mathrm{eff}}\Omega_{\mathrm{m}}}}.
\end{equation}
Here \(\Omega_{\mathrm{m}}=2\pi f_{\mathrm{m}}\), and the mechanical displacement field \(\mathbf{u}(\mathbf{r})\) is normalized as
\begin{equation}
\mathbf{Q}(\mathbf{r}) =
\frac{\mathbf{u}(\mathbf{r})}{D_{\max}},
\qquad
D_{\max}=\max |\mathbf{u}(\mathbf{r})|.
\end{equation}
The effective mass is then
\begin{equation}
m_{\mathrm{eff}} =
\rho_{\mathrm{Si}}
\int_{\mathrm{Si}}
|\mathbf{Q}(\mathbf{r})|^2
\,dV,
\end{equation}
with an additional symmetry factor included when only half of the mechanical structure is simulated.

\begin{figure*}[t]
\centering    
\includegraphics[width= 0.8\linewidth]{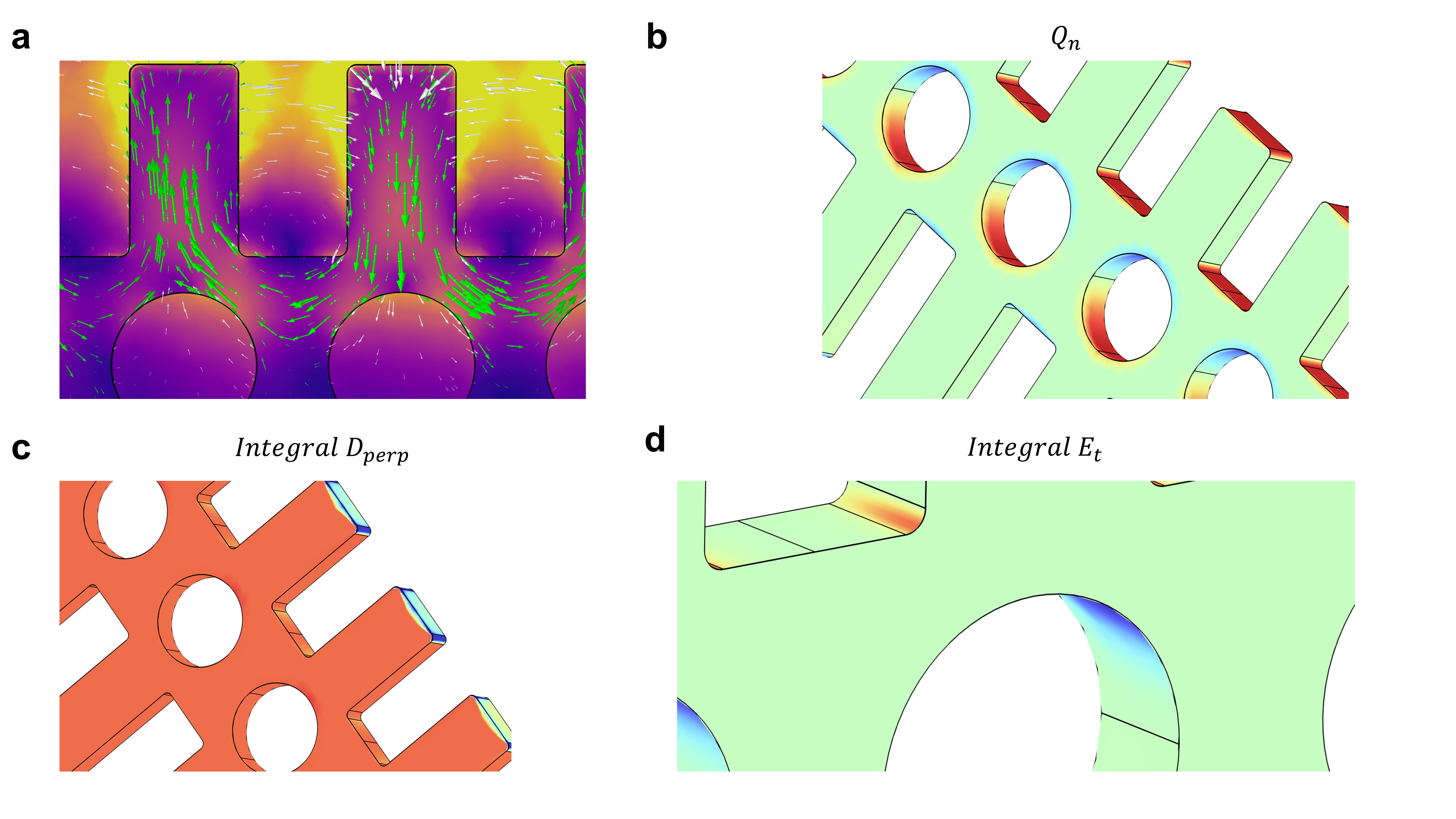}
\caption{ \textbf{a} Vectorial electric field (white arrows) and electric displacement field \(\mathbf{D}\) (green arrows), together with the normalized electric-field distribution. 
\textbf{b} Surface distribution of \(\mathbf{Q}\cdot\mathbf{n}\) for the one-antinode in-plane flexural mode. Red indicates positive values, while blue indicates negative values. 
\textbf{c} Contribution to \(g_\mathrm{MB}\) from the normal component of the electric displacement field, \(D_\perp\). 
\textbf{d} Contribution to \(g_\mathrm{MB}\) from the tangential component of the electric field, \(E_\parallel\). \label{fig: asym}}
\end{figure*}

The total coupling is calculated as the sum of the moving-boundary and photoelastic contributions,
\begin{equation}
\frac{g_0}{2\pi}
=
\frac{g_{0,\mathrm{MB}}}{2\pi}
+
\frac{g_{0,\mathrm{PE}}}{2\pi}.
\end{equation}
The moving-boundary contribution is evaluated from the perturbative surface integral
\begin{equation}
\frac{g_{0,\mathrm{MB}}}{2\pi}
=
-\frac{x_{\mathrm{zpf}}\nu_{\mathrm{c}}}{2}
\frac{
\oint_{\partial V}
(\mathbf{Q}\cdot\mathbf{n})
\left[
\Delta\varepsilon |\mathbf{E}_{\parallel}|^2
-
\Delta(\varepsilon^{-1}) |\mathbf{D}_{\perp}|^2
\right]
dA
}{
\int \varepsilon |\mathbf{E}|^2 dV
}.
\end{equation}
Here \(\partial V\) denotes the silicon--air boundary, \(\mathbf{n}\) is the outward normal to the silicon surface, \(\mathbf{E}_{\parallel}\) is the electric-field component parallel to the boundary, and \(\mathbf{D}_{\perp}\) is the electric-displacement component normal to the boundary. The dielectric discontinuities are defined as
\(\Delta\varepsilon=\varepsilon_{\mathrm{Si}}-\varepsilon_{\mathrm{air}}\) and
\(\Delta(\varepsilon^{-1})=\varepsilon_{\mathrm{Si}}^{-1}-\varepsilon_{\mathrm{air}}^{-1}\).

The photoelastic contribution is computed from the volume perturbation inside silicon,
\begin{equation}
\frac{g_{0,\mathrm{PE}}}{2\pi}
=
\frac{x_{\mathrm{zpf}}\nu_{\mathrm{c}}}{2}
\frac{
\int_{\mathrm{Si}}
\varepsilon_0 n_{\mathrm{Si}}^4
\mathbf{E}^{*}\cdot
\mathbf{p}:\mathbf{S}_{\mathbf{Q}}
\cdot\mathbf{E}
\,dV
}{
\int \varepsilon |\mathbf{E}|^2 dV
}.
\end{equation}
Here \(n_{\mathrm{Si}}\) is the refractive index of silicon, \(\mathbf{p}\) is the photoelastic tensor, and \(\mathbf{S}_{\mathbf{Q}}\) is the strain tensor associated with the normalized displacement field \(\mathbf{Q}\). The denominator in both perturbative expressions is the electromagnetic energy normalization integral.

For the flexural modes studied here, the optomechanical coupling is dominated by the moving-boundary term, while the photoelastic contribution is comparatively smaller. If the optical mode is perfectly symmetric along the transverse direction of the nanobeam, in-plane flexural modes couple inefficiently due to symmetry. Therefore, to control and enhance the coupling, we introduce a transverse asymmetry in the optical defect, as explained in the main text. This shifts the optical field towards one side of the nanobeam, increasing its spatial overlap with the mechanical displacement field.

Although the coupling increases with this asymmetry, it does not reach very large values. In particular, the fundamental flexural mode still exhibits a lower coupling rate than higher-order flexural modes. This can be understood by separating the two terms entering the moving-boundary integral in Eq.~(6): the electric field parallel to the boundary, $E_{\parallel}$, and the electric displacement field perpendicular to it, $D_{\perp}$. Figure~\ref{fig: asym} shows these vector fields together with the normalized electric-field distribution.

The regions that contribute most strongly are those where either $E_{\parallel}$ or $D_{\perp}$ is large at the dielectric boundary. The sign of each local contribution is determined by the normal component of the mechanical displacement, as shown in Fig.~\ref{fig: asym}b. Figures~\ref{fig: asym}c,d show the resulting spatial maps for this particular mode. The dominant regions are located at the outer boundaries, the wings, and the inner part of the curved sections, even though the corners are filleted to avoid singularities.

For the case $h_2/h_1 = 1.32$, the term associated with $E_{\parallel}$ is very large, giving $g_{0,\mathrm{MB},E_{\parallel}}/2\pi \approx 1.33~\mathrm{MHz}$. However, the $D_{\perp}$ term has the opposite sign and almost cancels it, even though most of the optical power is concentrated in the same region. As a result, the net coupling of the fundamental flexural mode is relatively small and sensitive to small geometrical variations. To account for these numerical fluctuations, we perform three simulations with slightly different values of $h_2/h_1$ around each nominal point.

The third- and fifth-order in-plane flexural modes also exhibit partial cancellation between the $E_{\parallel}$ and $D_{\perp}$ terms, but to a lesser extent. Although their individual moving-boundary terms are smaller, the cancellation is less pronounced, resulting in larger net optomechanical coupling rates.

Selecting different optical modes or exploring alternative asymmetry designs could reduce this cancellation and bring the coupling closer to the MHz regime in a non-slot cavity.

\subsection*{Experimental measurement}

As described in the main text, we follow an approach similar to Ref.~\cite{Gorodetsky2010OE23236}, using a phase-modulation tone to calibrate the vacuum optomechanical coupling rate. The phase modulation generates a calibration tone with known modulation depth
\begin{equation}
\beta = \frac{\pi V_{\mathrm{pk}}}{V_{\pi}},
\end{equation}
where $V_{\mathrm{pk}}$ is the peak voltage applied to the phase modulator and $V_{\pi}$ is its half-wave voltage at the calibration frequency. By comparing the integrated area of the mechanical peak, $A_{\mathrm{m}}$, with that of the calibration tone, $A_{\mathrm{cal}}$, the vacuum optomechanical coupling rate is obtained as
\begin{equation}
g_0 \approx
\sqrt{
\frac{A_\mathrm{m}}{A_\mathrm{cal}}
\frac{\beta^2 \Omega_\mathrm{cal}^2}{4 n_\mathrm{th}}
}.
\end{equation}
This expression assumes that $n_{\mathrm{th}}\gg 1$ and that the calibration tone and the mechanical signal experience the same transduction. Here, $\Omega_{\mathrm{cal}}$ is the angular frequency of the calibration tone and $n_{\mathrm{th}}$ is the thermal occupation of the mechanical mode,
\begin{equation}
n_{\mathrm{th}} =
\frac{1}{\exp\left(\hbar\Omega_{\mathrm{m}}/k_{\mathrm{B}}T\right)-1}
\simeq
\frac{k_{\mathrm{B}}T}{\hbar\Omega_{\mathrm{m}}}.
\end{equation}

At MHz frequencies, however, we find that our phase modulator exhibits dominant residual amplitude modulation (RAM), which overwhelms the phase-to-amplitude transduction produced by the cavity. We therefore place the calibration tone at a higher frequency, where RAM is strongly reduced. This procedure is valid provided that the detection chain has an equivalent response in both frequency ranges, as is the case here because the fast photodetector exhibits a flat response over the relevant bandwidth.

\begin{figure*}[t]
\centering    
\includegraphics[width= 0.8\linewidth]{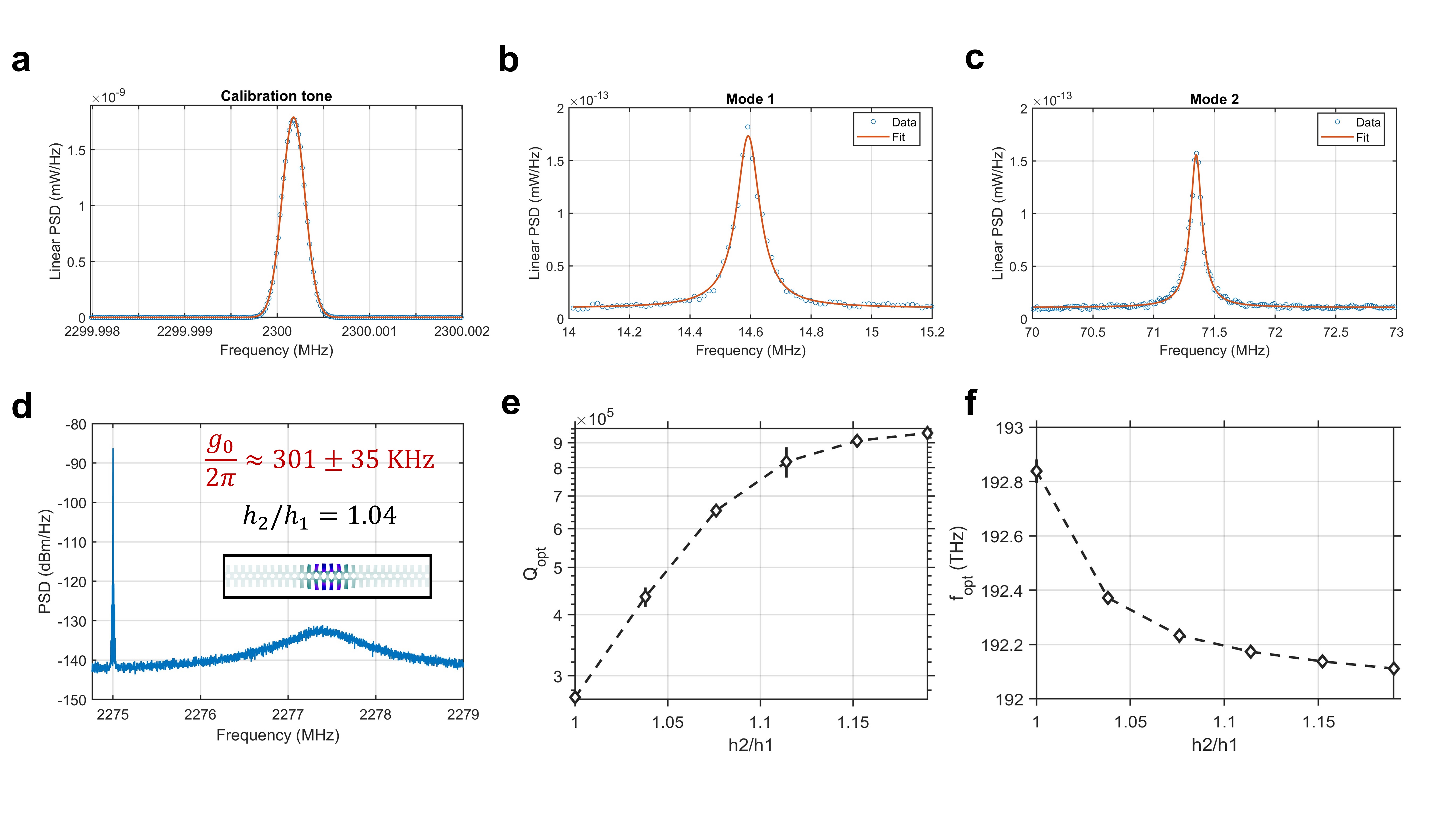}
\caption{ \textbf{a--c} Linear power spectral density measured with the signal analyzer in the direct-detection scheme for the calibration tone and the one- and three-antinode flexural modes. \textbf{d} RF signal of the calibration tone near a high-frequency breathing mode. The inset shows the displacement field of this mechanical mode. \textbf{e,f} Simulated quality factor and frequency of the optical mode as a function of the asymmetry.\label{fig: optmec}}
\end{figure*}

Figures~\ref{fig: optmec}a--c show the corresponding calibration tone, placed around $2.3~\mathrm{GHz}$, together with the first two in-plane flexural modes with an antinode at the cavity center. The calibration tone is fitted with a Gaussian function, whereas the mechanical modes are fitted with Lorentzian functions.

We further validate this calibration procedure using a high-frequency breathing mode, for which simulations predict a coupling rate of approximately $315~\mathrm{kHz}$ for $h_2/h_1 = 1.04$. We measure three different devices with this ratio and obtain $g_0/2\pi = 301 \pm 35~\mathrm{kHz}$, in good agreement with the simulated value.

Finally, we show the simulated dependence of the optical quality factor and resonance frequency on the cavity asymmetry. The simulated quality factor increases from $3\times10^5$ to almost $10^6$ as the asymmetry is increased. In practice, however, the intrinsic quality factor of the measured devices saturates around $10^5$, likely due to surface roughness, and we therefore do not observe a strong dependence on the asymmetry. On the other hand, the optical resonance frequency increases slightly as the cavity becomes more symmetric, which is not significant for the analysis presented here.

\section*{S5 Band structure of serpentine interconnects}

In the main text, the parameter \(s\) is used to tune the inter-cavity coupling by shifting the relevant band edge of the serpentine interconnect. Here, we complement that analysis by comparing the effect of the three geometrical parameters \(R\),
\(s\), and \(t\) on the stop-band regions of the unit cell.

\begin{figure*}[t]
\centering    
\includegraphics[width= 0.8\linewidth]{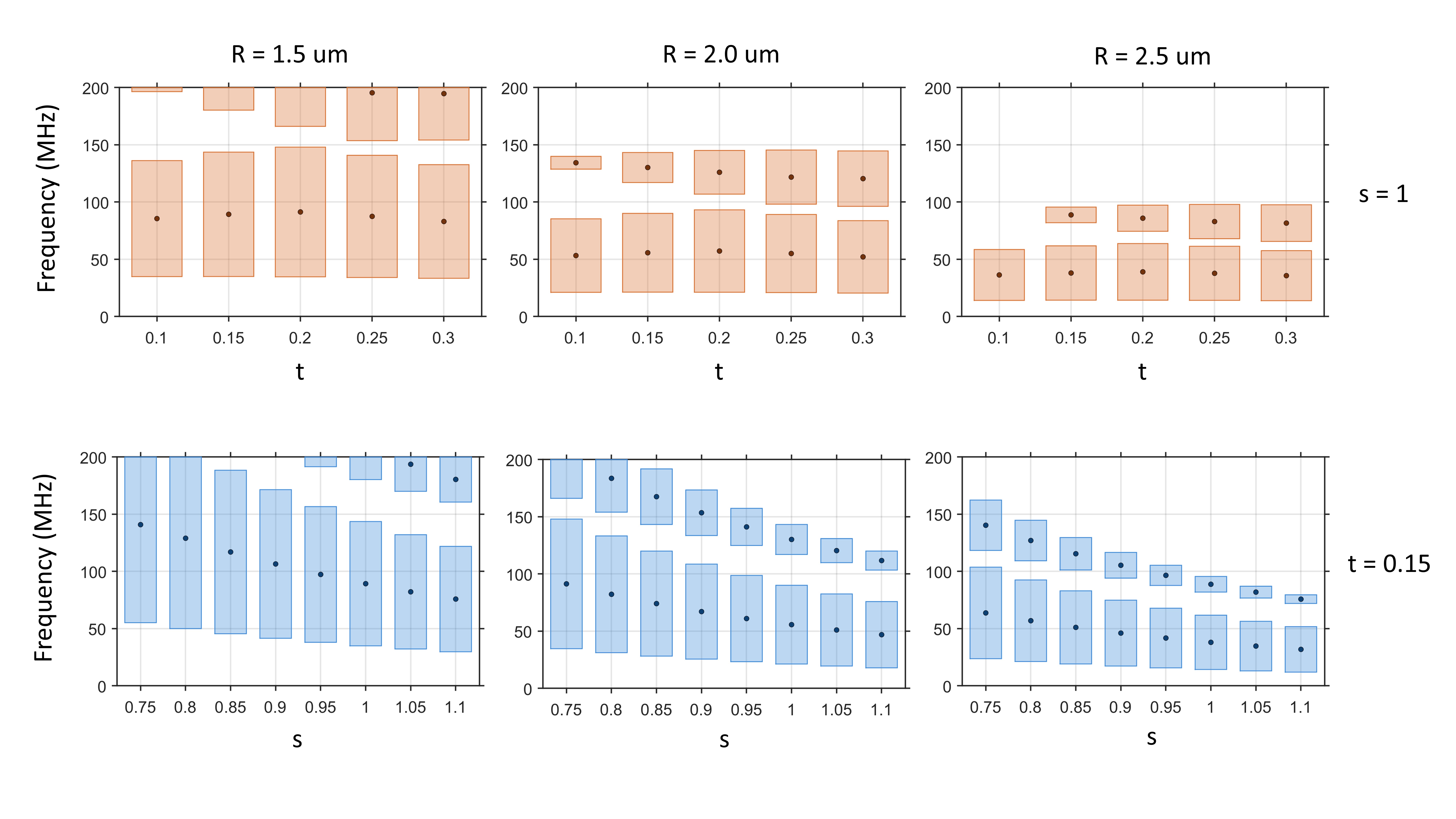}
\caption{ Stop-band frequency regions of the serpentine unit cell as a function of the geometrical parameters \(R\), \(t\), and \(s\). When \(t\) is swept, \(s=1\); when \(s\) is swept, \(t=0.15\). The points indicate the center frequencies of
the gaps. The sweep shows that \(R\) mainly sets the mechanical frequency scale, \(t\) trades compactness against gap opening, and \(s\) provides a controlled band-edge shift used for coupling tuning in the main text. \label{fig: gaps}}
\end{figure*}

Figure~\ref{fig: gaps} shows the stop-band frequency ranges obtained from Bloch-periodic simulations. Decreasing \(R\) shifts the gaps to higher frequencies and generally broadens them, reflecting the reduction of the overall mechanical
length scale. The parameter \(t\), which controls the horizontal extent of the ellipse and therefore the projected pitch of the interconnect, mainly affects the compactness of the structure. Reducing \(t\) makes the unit cell shorter, which is favourable for compact devices, but it also tends to close the relevant gaps, limiting its usefulness as a coupling-tuning parameter.

By contrast, varying \(s\) shifts the band edges in a controlled way while preserving the footprint and without strongly compromising the gap opening over the range used experimentally. This makes \(s\) the most convenient parameter for
tuning the position of the confined flexural mode relative to the stop-band edge, and therefore for controlling the evanescent attenuation discussed in the main text.

\section*{S6 Simulation of the mechanical quality factor}

\begin{figure*}[t]
\centering    
\includegraphics[width= 0.8 \linewidth]{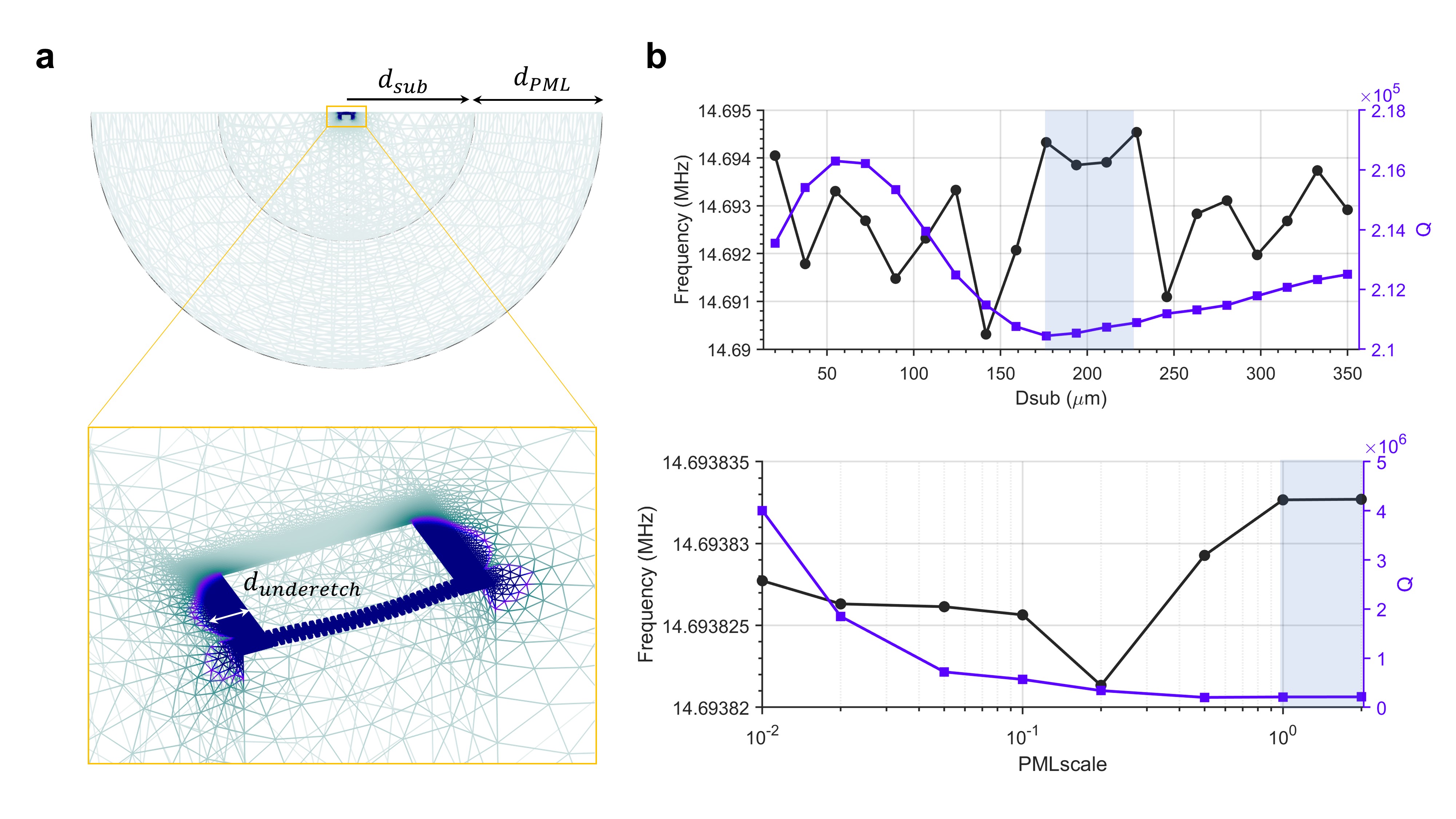}
\caption{
\textbf{a} FEM simulation geometry used to estimate the mechanical anchor-loss-limited quality factor. The nanobeam is connected to a suspended membrane region that emulates the silicon left by the underetching process. The membrane is attached to a quarter-spherical silicon substrate of radius \(D_\mathrm{sub}\), surrounded by a PML shell of thickness \(D_\mathrm{PML}\), which absorbs outgoing elastic waves radiated into the substrate.
\textbf{b} Simulated eigenfrequency and mechanical quality factor of the one-antinode in-plane flexural mode as a function of substrate radius \(D_\mathrm{sub}\), for fixed \(D_\mathrm{PML}=200~\mu\mathrm{m}\), and as a function of the PML scaling parameter, for fixed \(D_\mathrm{sub}=200~\mu\mathrm{m}\) and \(D_\mathrm{PML}=200~\mu\mathrm{m}\). The shaded blue regions indicate the selected operating regions used for the final simulations.
\label{fig:quality}
}
\end{figure*}

We estimate the mechanical quality factor of the nanobeams as a function of the number of unit cells using finite-element-method eigenfrequency simulations with perfectly matched layer (PML) domains. This approach is commonly used to estimate anchor-loss-limited mechanical dissipation~\cite{Bindel2005ElasticPMLs}. The mechanical quality factor is extracted from the complex mechanical eigenfrequency \(\tilde{f}_\mathrm{m}\) as
\begin{equation}
    Q_\mathrm{m} = \left| \frac{\mathrm{Re}(\tilde{f}_\mathrm{m})}{2\,\mathrm{Im}(\tilde{f}_\mathrm{m})} \right|,
\end{equation}
 The imaginary part accounts for elastic energy leaking from the resonator into the surrounding substrate.

It is worth noting that, at frequencies of a few tens of MHz, the wavelength of acoustic waves in bulk silicon is on the order of several hundreds of micrometres. Therefore, the simulation domain must be sufficiently large to avoid artificial reflections from the outer boundaries. We implement the model in COMSOL Multiphysics, as shown in Fig.~\ref{fig:quality}a. The nanobeam is connected to a suspended membrane region, which emulates the silicon left after the underetching process. This membrane is then attached to a silicon substrate represented by a quarter sphere of radius \(D_\mathrm{sub}\). A spherical PML shell of thickness \(D_\mathrm{PML}\) is added around the substrate to absorb outgoing elastic waves.

In practice, including the suspended membrane between the nanobeam and the bulk substrate is important to avoid unrealistically high simulated quality factors. We set the underetched membrane length to \(d_\mathrm{underetch} \approx 2~\mu\mathrm{m}\), which is consistent with the expected fabrication geometry. The substrate and PML are meshed using a swept mesh in the radial direction, with a characteristic resolution of \(10~\mu\mathrm{m}\).

We then analyze the convergence of both the mechanical frequency and quality factor as a function of the substrate radius and PML parameters, following a similar procedure to that described in Ref.~\cite{Li2022PMLAnchorLossCOMSOL}. First, we fix \(D_\mathrm{PML} = 200~\mu\mathrm{m}\) and sweep \(D_\mathrm{sub}\) from \(10~\mu\mathrm{m}\) to \(350~\mu\mathrm{m}\). As shown in Fig.~\ref{fig:quality}b, both the eigenfrequency and the quality factor show only a weak dependence on \(D_\mathrm{sub}\) over the explored range. We therefore choose \(D_\mathrm{sub} = 200~\mu\mathrm{m}\), which lies within a region where both quantities are stable.

Next, keeping \(D_\mathrm{sub}=200~\mu\mathrm{m}\) and \(D_\mathrm{PML}=200~\mu\mathrm{m}\) fixed, we sweep the PML scaling parameter. In contrast to the substrate radius, the extracted quality factor shows a stronger dependence on the PML scaling. For small values of the PML scaling parameter, the quality factor is artificially increased, indicating insufficient absorption of the outgoing elastic waves. For values larger than approximately unity, both the frequency and the quality factor reach a stable region. We therefore select an operating point in this convergence plateau, using \(D_\mathrm{sub}=200~\mu\mathrm{m}\), \(D_\mathrm{PML}=200~\mu\mathrm{m}\), and a PML scaling factor close to unity. This yields an anchor-loss-limited mechanical quality factor of approximately \(Q_\mathrm{m, clamp} \sim 1.6\times10^{5}\) for the simulated one-antinode in-plane flexural mode. 

\begin{figure*}[t]
\centering
\includegraphics[width=0.8\linewidth]{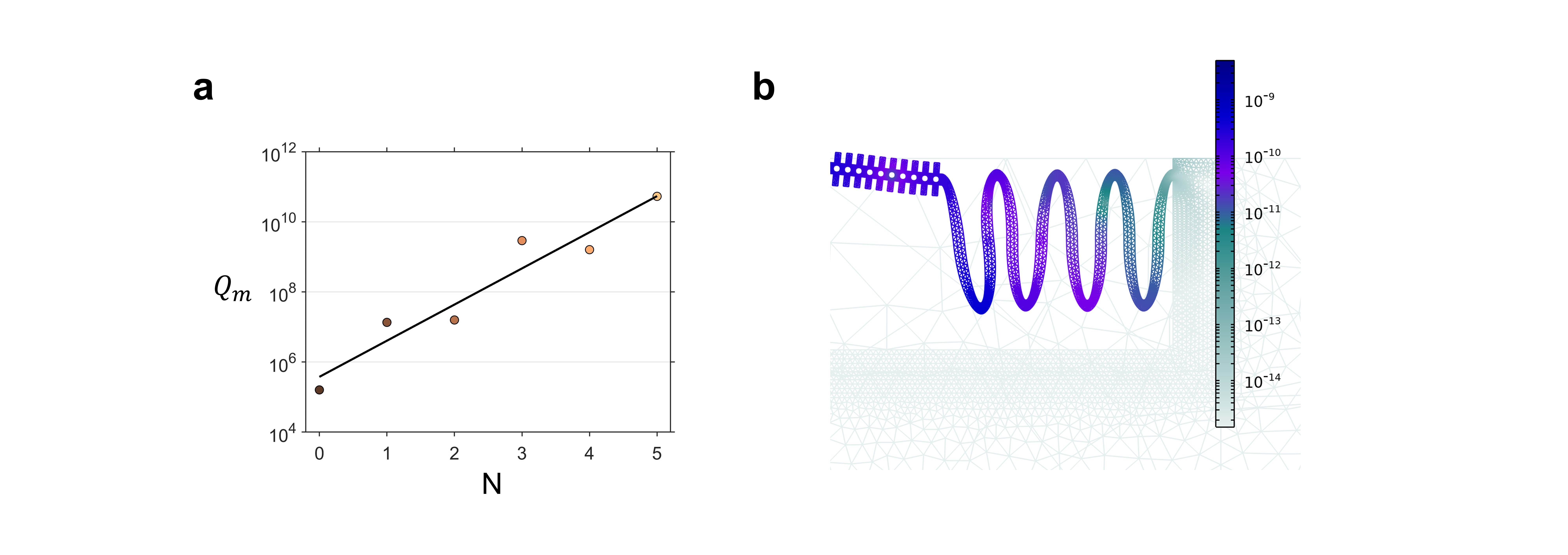}
\caption{\textbf{a} Simulated mechanical quality factor as a function of the number of serpentine clamping cells \(N\). A linear fit is performed on the logarithm of the data. \textbf{b} Displacement field on a logarithmic scale in the serpentine clamping region. \label{fig: dependenceN}}
\end{figure*}

Finally, we study the quality factor \(Q\) as a function of the number of serpentine unit cells connecting the resonator to the membrane. Figure~\ref{fig: dependenceN}a shows the results of this analysis. Despite an even--odd effect in the number of cells, likely related to the position at which the serpentine is clamped to the membrane, an exponential increase of \(Q\) is observed. Indeed, fitting the data as \(\log(Q)=\log(Q_0)+bN\) yields \(b=2.4\pm0.3\), corresponding to an attenuation of \(\alpha=10.3\pm1.5~\mathrm{dB/cell}\). This value is comparable to that obtained from the complex-band-structure analysis of the unit cell.

Figure~\ref{fig: dependenceN}b shows the displacement field on a logarithmic scale, illustrating leakage of the mode through the serpentine structure. A clear exponential attenuation is observed from cell to cell.

\section*{S7 Dependence on temperature and pressure}

Here we extend the pressure-dependent characterization discussed in the main text and also report the temperature dependence of the mechanical modes. 
\subsection*{Pressure analysis}

\begin{figure*}[t]
\centering
\includegraphics[width=0.8\linewidth]{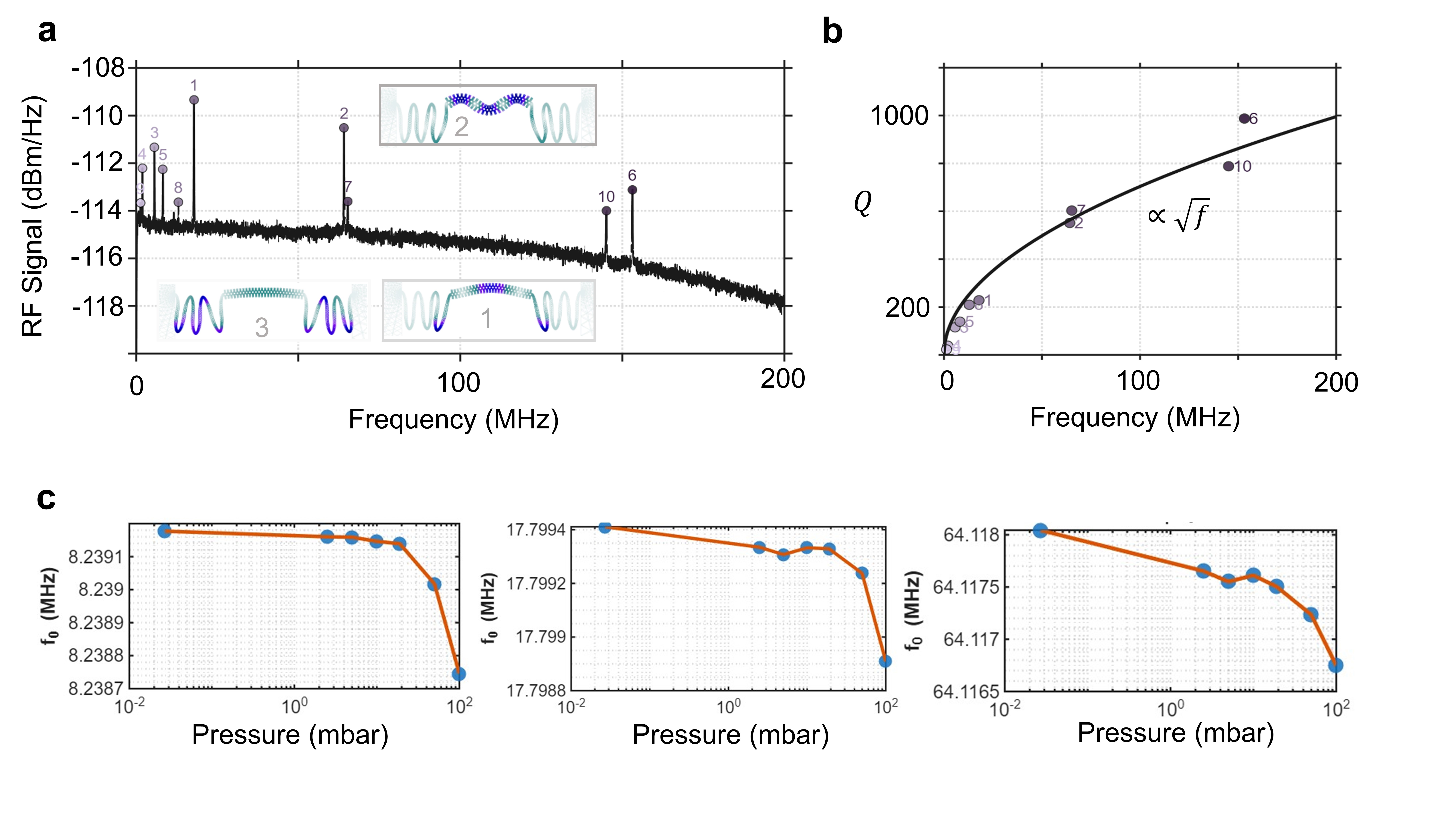}
\caption{
\textbf{a} RF spectrum of the mechanical modes of a serpentine-clamped single nanobeam with parameters \((N,r,s,t)=(3,2.5~\mu\mathrm{m},1,0.15)\), measured in
a grating--grating transmission configuration at atmospheric pressure. The inset shows the simulated displacement-field distributions of modes 1, 2, and 3. \textbf{b} Quality factor of the mechanical modes at atmospheric pressure as a function of resonance frequency. The dashed line shows a fit proportional to
\(\sqrt{f}\), consistent with gas damping in the viscous-flow regime.
}\label{fig: pressure}
\end{figure*}

Figure~\ref{fig: pressure}a shows the RF spectrum measured from a serpentine-clamped single nanobeam in a grating--grating transmission configuration, using direct detection and probing the fundamental optical cavity mode. In this spectrum, we identify ten mechanical resonances, from which we extract the resonance frequencies and quality factors. The resulting quality factors at atmospheric pressure are shown in Fig.~\ref{fig: pressure}b. The measured quality factors increase approximately as \(\sqrt{f_m}\), where \(f_m=\Omega_m/2\pi\). This trend is consistent with viscous gas damping, for which the relevant hydrodynamic length scale is the oscillatory boundary-layer thickness~\cite{Sader1998,Zengerle2022},
\begin{equation}
    \delta_{\rm vis}=\sqrt{\frac{2\mu}{\rho_{\rm gas}\Omega_m}}=\sqrt{\frac{\mu}{\pi\rho_{\rm gas}f_m}},
\end{equation}
with \(\mu\) and \(\rho_{\rm gas}\) the dynamic viscosity and mass density of the gas, respectively. Since \(\delta_{\rm vis}\propto f_m^{-1/2}\), the effective gas damping of comparable modes scales approximately as \(b_{\rm gas}\propto\sqrt{\Omega_m}\), giving
\begin{equation}
    Q_{\rm gas}=\frac{m_{\rm eff}\Omega_m}{b_{\rm gas}}\propto\sqrt{\Omega_m}\propto\sqrt{f_m}.
\end{equation}
The black line in Fig.~\ref{fig: pressure}b shows a fit proportional to \(\sqrt{f_m}\). This agreement should be interpreted as a qualitative indication that the modes are predominantly limited by air damping at atmospheric pressure, since the different resonances have different displacement profiles and overlaps with the surrounding gas.

For completeness, we also show the pressure dependence of the mechanical frequency for the three modes discussed in the main text. In contrast to the quality factor, which changes by almost two orders of magnitude over the measured pressure range, the resonance-frequency shifts are modest. This is consistent with gas damping primarily affecting the linewidth, while the reactive component of the fluid load produces only a modest resonance-frequency shift through added inertia~\cite{Sader1998}. These effects are strongest near atmospheric pressure, where the gas density is largest. As the pressure is reduced, the gas contribution rapidly becomes small compared with the intrinsic inertia and stiffness of the resonator, so further pressure reductions produce only weak additional frequency shifts.

\subsection*{Temperature analysis}
\begin{figure*}[t]
\centering
\includegraphics[width=0.8\linewidth]{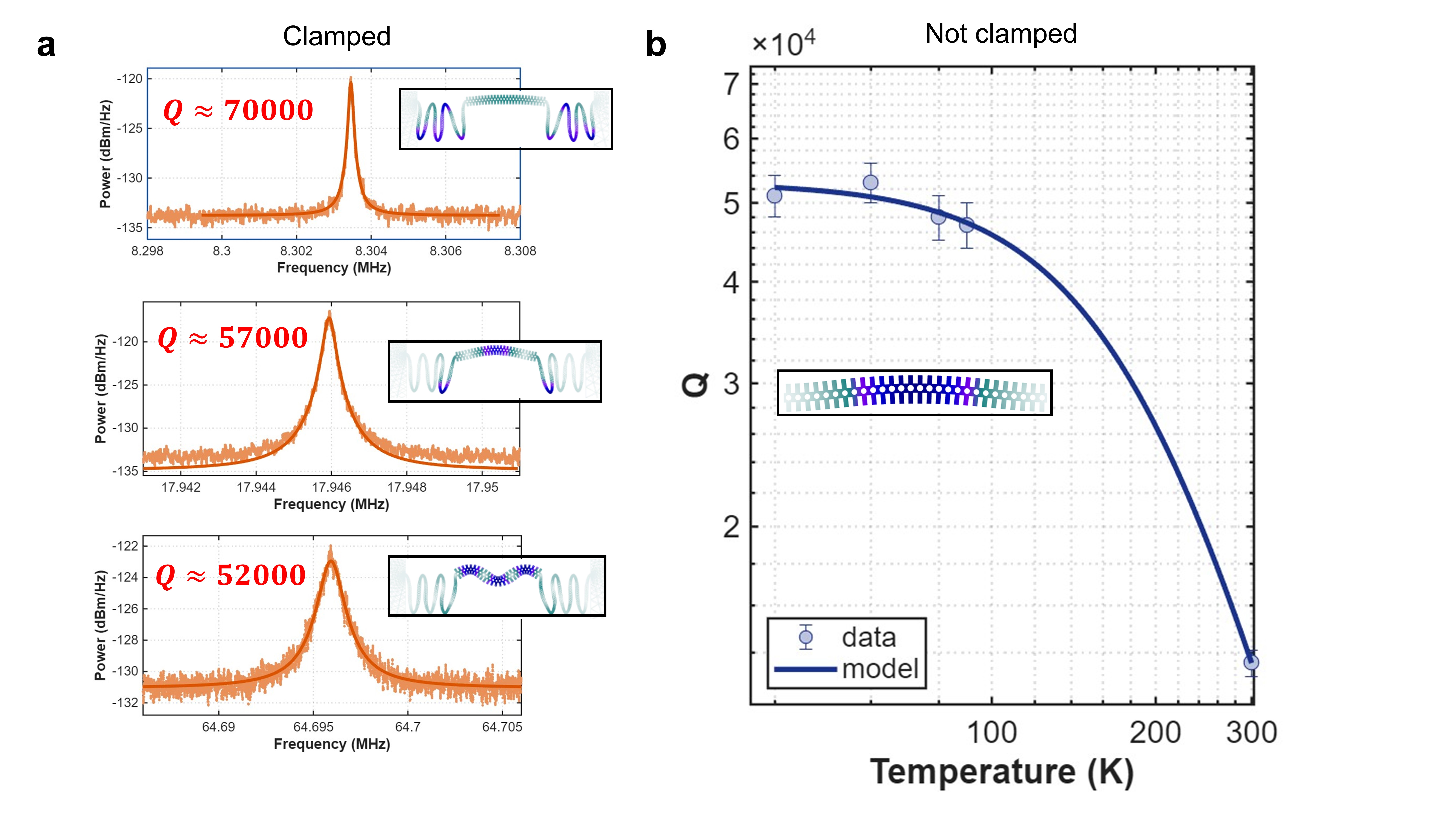}
\caption{
\textbf{a} RF spectra of the modes analyzed in the pressure-dependent measurements of the main text, measured at \(4~\mathrm{K}\) under high-vacuum conditions.
\textbf{b} Quality factor as a function of temperature for the fundamental in-plane flexural mode of a nanobeam without serpentine terminations. The solid line shows a saturation fit to the data.
}
\label{fig: temp}
\end{figure*}

At high vacuum, we further reduce the temperature to \(4~\mathrm{K}\) to assess whether the room-temperature quality factors of the serpentine-clamped single beams are limited by temperature-dependent dissipation. The quality factors of all the modes discussed in the main text increase considerably at low temperature, as shown in Fig.~\ref{fig: temp}a. This indicates that the room-temperature saturation is likely dominated by internal dissipation mechanisms, with thermoelastic damping being a possible contribution.

We also show in Fig.~\ref{fig: temp}b the temperature dependence of the quality factor of the fundamental in-plane flexural mode in a nanobeam without serpentine terminations. The quality factor saturates at a value comparable to that of the serpentine-clamped structure. This suggests that, consistently with the clamping-loss simulations of Section~S6, the clamping losses of this mode are already sufficiently low in the unclamped geometry, so that the additional stop-band isolation provided by the serpentine unit cells does not lead to an observable increase of \(Q\) in the present devices.

\section*{S8 Two-cavity readout and detuning correction}

We model the two coupled mechanical resonators with the effective Hamiltonian
\begin{equation}
H=
\begin{pmatrix}
f_1 & g\\
g & f_2
\end{pmatrix},
\end{equation}
where \(f_1\) and \(f_2\) are the uncoupled resonance frequencies and \(g\) is the coupling rate. The hybridized eigenfrequencies are
\begin{equation}
f_{\pm}=\frac{f_1+f_2}{2}\pm\frac{1}{2}\sqrt{\Delta^2+4g^2},
\qquad
\Delta=f_1-f_2,
\end{equation}
so that the measured splitting is
\begin{equation}
\Omega=f_+-f_-=\sqrt{\Delta^2+4g^2}.
\end{equation}
Only when the residual detuning is negligible compared with the coupling-induced splitting does \(\Omega\simeq 2g\). For finite detuning, using \(g_{\mathrm{app}}=\Omega/2\) slightly overestimates the true coupling, which is instead
\begin{equation}
g=\sqrt{g_{\mathrm{app}}^2-\left(\frac{\Delta}{2}\right)^2}.
\end{equation}

The same model gives the weight of each hybridized mode on the two local nanobeams,
\begin{equation}
w_{L,\pm}=\frac{1}{2}\left(1\pm\frac{\Delta}{\sqrt{\Delta^2+4g^2}}\right),
\qquad
w_{R,\pm}=\frac{1}{2}\left(1\mp\frac{\Delta}{\sqrt{\Delta^2+4g^2}}\right),
\end{equation}
up to the sign convention used to label \(f_1\) and \(f_2\). When \(|\Delta|\ll 2g\), both hybridized modes have significant weight on both cavities and can be observed in both local readout channels. When \(|\Delta|\) becomes comparable to \(2g\), the modes become partially localized and the two local spectra show different relative peak amplitudes.

\begin{figure*}[t]
\centering
\includegraphics[width=0.8\linewidth]{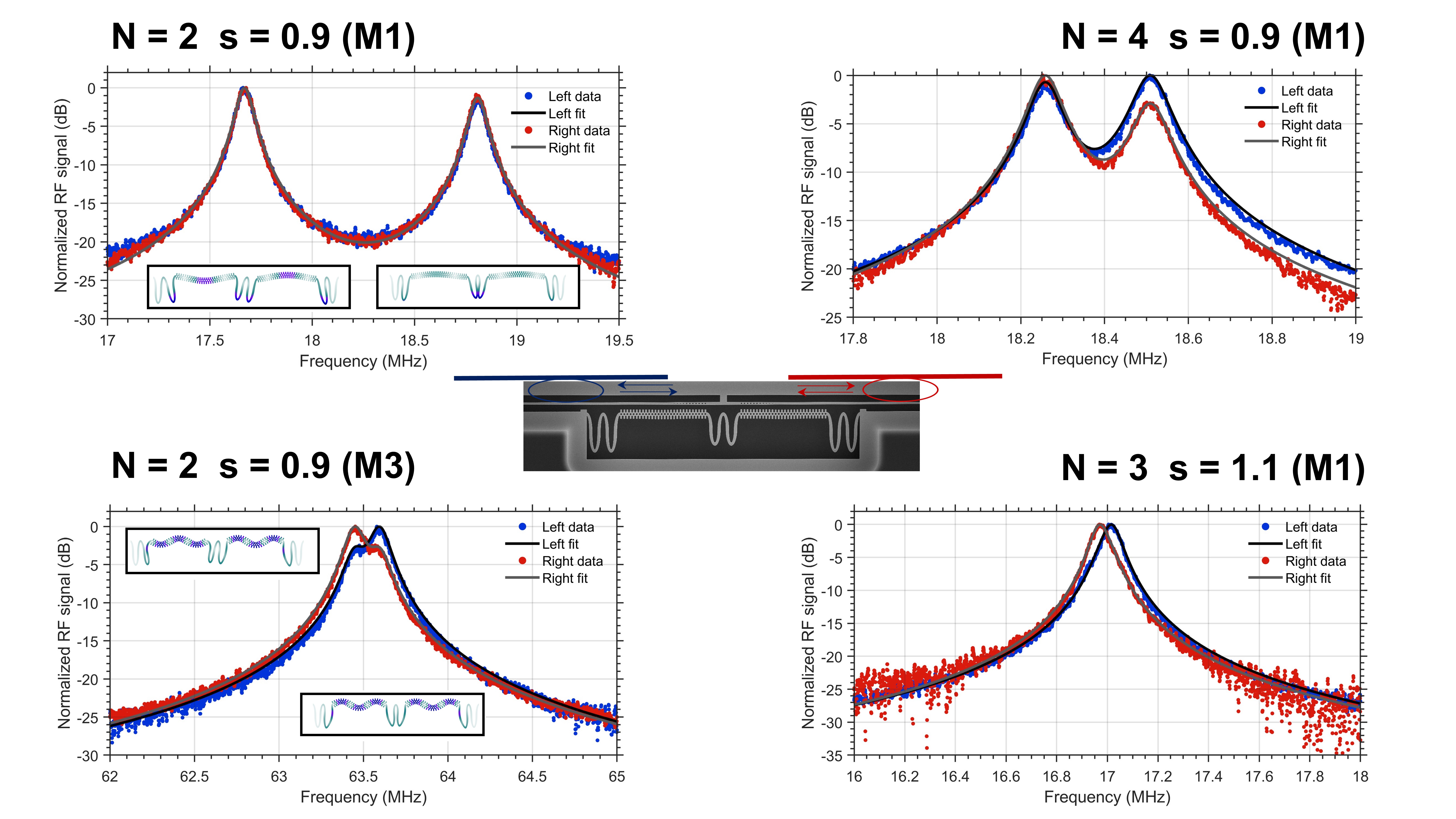}
\caption{
Two-cavity optomechanical readout of coupled flexural modes. The spectra measured from the left and right cavities are fitted simultaneously using a two-mode coupled-resonator model. The peak positions are constrained by the hybridized eigenfrequencies, while the relative peak amplitudes are determined by the eigenvector weights and a fixed optical visibility ratio \(r_\eta=\eta_-/\eta_+\) extracted from the strongly hybridized \(N=2\) (M1) configuration. Additional panels illustrate cases where the residual detuning becomes comparable to, or larger than, the coupling-induced splitting, leading to partially localized modes and different relative peak amplitudes in the two local readout channels.
}
\label{fig:double}
\end{figure*}

Figure~\ref{fig:double} shows representative two-cavity readout spectra. For the \(N=2\), \(s=0.9\) M1 configuration, both hybridized modes are clearly visible in both cavities with comparable relative amplitudes. This does not imply that the two bare resonators are perfectly frequency matched; rather, the coupling-induced splitting is large compared with the residual detuning. In this regime, the modes remain strongly hybridized and the approximation \(g\simeq\Omega/2\) is well justified.

However, the two hybridized modes do not have identical optical visibility. Experimentally, the antisymmetric mode is transduced more strongly than the symmetric mode, with \(A_{\mathrm{asym}}\simeq1.34A_{\mathrm{sym}}\). This is consistent with simulations, where the antisymmetric mode has a larger displacement amplitude in the nanobeams, while the symmetric mode is more concentrated around the central serpentine link. The symmetric mode therefore has a smaller optomechanical overlap with the optical cavity mode. We account for this effect through a fixed relative visibility factor
\begin{equation}
r_\eta=\frac{\eta_-}{\eta_+}\simeq1.34,
\end{equation}
where \(\eta_-\) and \(\eta_+\) are the optical transduction efficiencies of the antisymmetric and symmetric hybridized modes, respectively. This factor is extracted from the \(N=2\) configuration and kept fixed when fitting the \(N=4\) spectra, where the smaller coupling makes the residual detuning more apparent in the relative peak amplitudes measured from the two cavities.

Each hybridized resonance is modeled as a unit-height Lorentzian,
\begin{equation}
L_{\pm}(f)=\frac{\gamma_{\pm}^2}{(f-f_{\pm})^2+\gamma_{\pm}^2},
\end{equation}
and the spectra measured from the left and right cavities are fitted simultaneously as
\begin{align}
S_L(f)&=B_L+C_L\left[r_\eta w_{L,-}L_-(f)+w_{L,+}L_+(f)\right],\\
S_R(f)&=B_R+C_R\left[r_\eta w_{R,-}L_-(f)+w_{R,+}L_+(f)\right].
\end{align}
Here \(B_{L,R}\) and \(C_{L,R}\) account for the RF background and the overall readout efficiency of each cavity. The peak positions are constrained by the coupled-mode eigenfrequencies, while the relative peak amplitudes are constrained by the eigenvector weights and corrected by the independently calibrated visibility factor \(r_\eta\).

For the \(N=4\), \(s=0.9\) M1 device, the two local spectra are nearly mirrored, indicating a finite residual detuning. The fit gives modal weights of approximately \(0.61\) and \(0.39\), corresponding to \(|\Delta|\simeq55~\mathrm{kHz}\) and \(2g\simeq244~\mathrm{kHz}\). Thus, even in this case, \(|\Delta|<2g\), and both modes remain significantly hybridized. The apparent coupling obtained directly from the measured splitting is \(g_{\mathrm{app}}\simeq125~\mathrm{kHz}\), while the detuning-corrected value is \(g\simeq122~\mathrm{kHz}\). The correction is therefore only about \(3~\mathrm{kHz}\), confirming that using half of the measured splitting gives a reliable estimate of the coupling in the regime analyzed in the main text.

It is also worth noting why, in the main text, we extract the coupling decay with \(N\) only for M1, i.e., the one-antinode in-plane flexural mode. As shown in Figs.~2c,d of the main text, the attenuation per serpentine cell is considerably larger for M3, the three-antinode in-plane flexural mode. As a result, already for \(N=2\), the expected splitting is below \(100~\mathrm{kHz}\), and the residual frequency mismatch between nominally identical cavities can become comparable to or larger than the coupling-induced splitting. In this regime, the local spectra are dominated by the intrinsic detuning rather than by hybridization, and both modes are no longer necessarily visible from a single optical cavity. Figure~\ref{fig:double} illustrates this effect with two representative cases. For M3 with \(N=2\), the spectra already resemble the partially localized situation observed for M1 with \(N=4\), \(s=0.9\). In this case, the two spectra are nearly mirror images because the symmetric and antisymmetric M3 modes have similar optical visibility. The figure also shows an example where the intrinsic detuning clearly dominates over the coupling-induced hybridization, corresponding to M1 with \(N=3\), \(s=1.1\).

\section{References}

%

\end{document}